\documentclass[showpacs,floatfix,pra,twocolumn,10pt]{revtex4}
\usepackage{amssymb}
\usepackage{graphicx}
\usepackage{amsmath,amssymb}
\usepackage{bm}

\begin{document}

\title{Robust fermionic-mode entanglement of a nanoelectronic system in non-Markovian environments}
\author{Jiong Cheng$^{1,}$\footnote{qu8key@mail.dlut.edu.cn},
Wen-Zhao Zhang$^{1}$, Yan Han$^{2,1}$ and Ling Zhou$^{1,}$\footnote{zhlhxn@dlut.edu.cn}}
\address{$^{1}$School of Physics and Optoelectronic Engineering, Dalian University of
Technology, Dalian 116024, China\\
$^{2}$School of Physics and Optoelectronic Technology, Taiyuan University of
Technology, Taiyuan 030024, China}
\date{\today}

\begin{abstract}

A maximal steady-state fermionic entanglement of a nanoelectronic system is generated in finite temperature
non-Markovian environments. The fermionic entanglement dynamics is presented by connecting the exact solution of the system
with an appropriate definition of fermionic entanglement. We prove that the two understandings of the dissipationless
non-Markovian dynamics, namely the bound state and the modified Laplace transformation are completely equivalent. For
comparison, the steady-state entanglement is also studied in the wide-band limit and Born-Markovian approximation. When the
environments have a finite band structure, we find that the system presents various kinds of relaxation processes. The
final states can be: thermal or thermal-like states, quantum memory states and oscillating quantum memory states.
Our study provide an analytical way to explore the non-Markovian entanglement dynamics of identical fermions in a realistic
setting, i.e., finite temperature reservoirs with a cutoff spectrum.

\end{abstract}
\maketitle

\section{\protect\bigskip Introduction}

Entanglement, a particularly striking feature of quantum mechanics is thought of as the key ingredients in the fields
of quantum information science \cite{Amico80 517,Horodecki81 865}. The theoretical study of the entanglement evolution in
open quantum systems has attracted considerable interest \cite{Zyczkowski65 012101,Yu93 140404,Sinaysky78 062301}. Recently,
much attention has been paid to the case when the non-Markovian effect is non-ignorable  \cite{Bellomo99 160502,
Wang15 103020,Benedetti87 052328,Xu104 100502}. The dynamical back action of the memory environments has been experimentally
observed \cite{Xu104 100502}, which provides a reliable way to preserve entangled states \cite{Bellomo99 160502,
Maniscalco100 090503,Yang87 022312,Tong81 052330}. The physical mechanism of this remarkable phenomenon
is understood as the interplay between the existence of the bound state and the non-Markovian effect \cite{Tong81 052330,
John50 1764,Lodahl430 654,Bellomo78 060302}. Most recently, Zhang \textit{et al}. explored the non-Markovian dynamics from
the modified Laplace transform \cite{Zhang109 170402}, which gives an accurate description of the non-Markovian memory effect.

As a candidate for realizing building blocks of quantum information processors, quantum dot (QD) nanostructures with tunability
of various couplings and energy levels is regarded as a promising quantum device \cite{Elzerman67 161308,Hayashi91 226804}.
Entanglement in such nanostructures has been studied extensively, ranging from spin entanglement
\cite{Recher63 165314,Legel76 085335,Erbe85 155127}, entangling an excitonic two-level system \cite{Blattmann89 012327} and
tripartite entanglement \cite{Posazhennikova88 042302,Hiltunen89 115322}, to entanglement detection or measurement
\cite{Blaauboer95 160402,Emary80 161309,Borras84 033301}. In practice, however, as the quantum system always interacts with
its environment, quantum decoherence and dissipation is usually the central impediment for maintaining the electron coherence.
Among the decoherence mechanisms, the non-Markovian effect of nanostructures plays an important role
\cite{Thorwart72 235320,Liang72 245328,Cao76 115301,Marcos83 125426,Madsen106 233601}. Since the non-Markovian entanglement
dynamics has been well understood in the distinguishable particle systems \cite{Tong81 052330,John50 1764,Lodahl430 654,
Bellomo78 060302,Zhang109 170402}, it is therefore natural to ask for an extension to the case of identical particle systems,
especially for nanoelectronic systems consisting of identical fermions. The corresponding non-Markovian fermionic entanglement
dynamics of these systems are still virtually unexplored.

To achieve a comprehensive understanding of the decoherence dynamics, one has to rely on realistic and precise model
calculations. Recently, exact master equations describing the non-Markovian dynamics for various nanodevices
\cite{Tu78 235311,Jin12 083013} have been developed by means of the Feynman-Vernon influence functional theory.
On the other hand, the standard definition of entanglement is no longer valid for identical particles due to the
conflicts between the indistinguishability of the system constituents and the tensor product structure \cite{Verch17 545} of
the Hilbert space. A multitude of early studies are focusing on entanglement between fixed numbers of indistinguishable
particles \cite{Schliemann63 085311,Schliemann64 022303,Paskauskas64 042310,Li64 054302,Eckert299 88,Shi67 024301,Wiseman91 097902}.
In general case, however, the particle numbers of an open system are not conserved, it is therefore reasonable to consider the
entanglement between fermionic modes, in a similar way as is conventionally done for bosonic modes \cite{Friis87 022338}.

In this paper, our main purpose is to present an analytical evaluation of fermionic entanglement in non-Markovian environments,
so as to seek a robust way to prepare entangled states. This is achieved by connecting the exact solution of the system with an appropriate definition of fermionic entanglement in the fermionic Fock space. To be specific, we consider a system of double quantum dots (DQDs) coupled to two electrodes. By utilizing the Grassmann calculus and the modified Laplace transformation, we find that the non-Markovian environments can lead to a maximal steady-state fermionic entanglement. Our analysis shows that the usual decoherence
suppression schemes implemented in distinguishable particle systems can also be achieved for identical fermions.

The rest of the article is structured as follows: In Sec. II we consider a general model to describe a two-mode fermion
hopping system subject to noninteracting fermionic environments. In Sec. III, a fully analytical approach of fermionic
entanglement dynamics is established. Then, in Sec. IV, some analytical and numerical results are discussed. Finally, we
discuss a more realistic scenario, namely finite temperature reservoirs with a cutoff spectrum, and conclude the paper in
Sec. V.

\section{Model Hamiltonian and the exact propagating function}

We consider a general two-mode fermion hopping system coupled to noninteracting fermionic environments. The total Hamiltonian
is given by
\begin{subequations}
\begin{eqnarray}
\hat{H}_s &=& \epsilon_1 \hat{a} _{1}^\dag \hat{a} _{1} + \epsilon_2 \hat{a} _{2}^\dag \hat{a} _{2}
            + G \hat{a} _{1}^\dag \hat{a} _{2} + G^* \hat{a} _{2}^\dag \hat{a} _{1}, \\
\hat{H}_E &=& \sum\limits_{lk} {\epsilon _{lk} \hat{b}_{lk}^\dag \hat{b}_{lk}}, \\
\hat{H}_I &=& \sum\limits_{jlk} {V_{jl}^{k*}\hat{a} _{j}^\dag\hat{b}_{lk}+V_{jl}^{k}\hat{b}_{lk}^\dag\hat{a} _{j}}.
\end{eqnarray}
\label{Htot}
\end{subequations}
The fermionic creation and annihilation operators $\hat{a} _{j}^\dag$ and $\hat{a} _{j}$ satisfy the canonical anticommutation relations (CARs), $\epsilon_j$ is the energy for the $j$-th mode, and $G$ is coupling strength between the two fermionic modes.
Similarly, $\hat{b}_{lk}$ and $\hat{b}_{lk}^\dag$ are the annihilation and creation operators of the $k$-th mode of two
fermionic environments ($l=1,2$) with the continuous energy $\epsilon _{lk}$. The coupling strength between the system and
the $k$-th mode of the environments is given by $V_{jl}^{k}$. Physically, such a system may be realized by DQDs in
which each dot has a single on-site energy level.

The open system (\ref{Htot}) can be solved analytically by using the Feynman-Vernon influence functional theory
\cite{Feynman24 118} in the fermionic coherent state representation \cite{Zhang62 867,Cahill59 1538}. The exact
propagating function was derived in \cite{Tu78 235311,Jin12 083013},
\begin{eqnarray}
\mathcal{J}(\eta'_f,\eta_{f}^{*};t|\eta'^{*}_{i},\eta_i;0)
&=&A(t)\exp\{ \eta_{f}^{*}J_{1}(t)\eta_{i}+\eta'^{*}_{i}J_{1}^{\dag}(t)\eta'_{f}  \notag\\
&&+\eta_{f}^{*}J_{2}(t)\eta'_{f}+\eta'^{*}_{i}J_{3}(t)\eta_{i}\},
\label{prop}
\end{eqnarray}
where
\begin{subequations}
\begin{eqnarray}
J_{1}(t)=\mathcal{W}(t)\mathcal{U}(t),~~~~~J_{2}(t)=\mathcal{W}(t)-\mathbf{I},~~~~ \\
J_{3}(t)=\mathcal{U}^{\dagger}(t)\mathcal{W}(t)\mathcal{U}(t)-\mathbf{I},~A(t)=\mathrm{det}^{-1}\mathcal{W}(t),
\end{eqnarray}
\label{coJ}
\end{subequations}
with $\mathcal{W}(t)=\frac{1}{\mathbf{I}-\mathcal{V}(t)}$ and $\mathbf{I}$ is an identity matrix.
$\eta_{i}$ and $\eta_{f}$ represent two sets of Grassmann variables, associated with initial and final fermionic
coherent states respectively \cite{Berezin1966,Cahill59 1538}. The time-dependent coefficients, namely $\mathcal{U}(t)$
and $\mathcal{V}(t)$, can be fully determined by the Dyson equation and the nonequilibrium fluctuation-dissipation
theorem respectively
\begin{subequations}
\begin{eqnarray}
\dot{\mathcal{U}}(\tau)+i \mathbf{M}\mathcal{U}(\tau)+\int_{0}^{\tau}d\tau'g(\tau-\tau')\mathcal{U}(\tau')=0,~~~ \label{u}\\
\mathcal{V}(t)=\int_{0}^{t}d\tau_1\int_{0}^{t}d\tau_2\mathcal{U}(t-\tau_1)\tilde{g}(\tau_1-\tau_2)\mathcal{U}^{\dagger}(t-\tau_2),
\label{v}
\end{eqnarray}
\label{uv}
\end{subequations}
subjected to the initial conditions $\mathcal{U}(0)=\mathbf{I}$ and $\mathcal{V}(0)=0$. The $2\times2$ matrices $\mathbf{M}=\left(\begin{array}{cc}\epsilon_{1} & G  \\ G^{*} & \epsilon_{2} \\ \end{array} \right)$ and the
non-local time correlation functions reads $g(\tau)=\sum\limits_{l}\int \frac{d\omega}{2\pi} J_{l}(\omega)e^{-i\omega\tau}$, $\tilde{g}(\tau)=\sum\limits_{l}\int\frac{d\omega}{2\pi}J_{l}
(\omega)\bar{n}_{l}(\omega,T)e^{-i\omega\tau}$. The spectral density $J_{l}(\omega)$ depends on the specific structure of the environment and the system-environment coupling strength $V_{jl}^{k}$. Furthermore, $\bar{n}_{l}(\omega,T)=\frac{1}{e^{\beta_{l}
(\omega-\mu_{l})}+1}$ is the initial Fermi-Dirac distribution of fermionic reservoir $l$ with chemical potential $\mu$ at
initial temperature $\beta=1/k_{B}T$.

The propagating function (\ref{prop}) together with the equations (\ref{uv}) totally determine the non-Markovian dynamics of
the open system (\ref{Htot}). In the following sections, by exactly solving the reduced density matrix
\begin{eqnarray}
\rho(\eta_{f}^{*},{\eta'_f};t)&=&\int d\mu({\mathbf{\eta}_i})d\mu({\mathbf{\eta'}_i})\mathcal{J}(\eta'_f,\eta_{f}^{*};t|\eta'^{*}_{i},\eta_i;0) \notag\\
&&\times\rho(\eta_{i}^{*},\eta'_{i};0),
\label{rdrho}
\end{eqnarray}
and analyzing the solution of nonequilibirum Green's functions $\mathcal{U}(t)$ and $\mathcal{V}(t)$ \cite{Zhang109 170402},
we will present an accurate way to take into account non-Markovian memory effects on the dynamics of fermionic entanglement.


\section{Exact decoherence dynamics}

\subsection{The exact solution of the system}

The explicit form of propagating function (\ref{prop}) lead to an exact master equation \cite{Tu78 235311,Jin12 083013,
Zhang109 170402}. However according to Grassmann calculus \cite{Berezin1966,Cahill59 1538}, Eq. (\ref{rdrho}) is exactly
solvable for arbitrary initial state $\rho(\eta_{i}^{*},\eta'_{i};0)$. This provides us a direct way to analyze the
evolution of the system. On the other hand, because of the superselection rules, the Q symbol (density matrix in the
fermionic coherent states representation) is restricted to even parity. To be specific, we should consider the case $\rho(\eta_{i}^{*},{\eta'_i};0)=\xi_{i}^{*}\rho_{1}\xi'_{i}+\eta_{i}^{*}\rho_{2}\eta'_{i}$, where the $2\times2$ matrix $\rho_{1}$ and $\rho_{2}$ include complete information of the initial states, while the two vectors are $\xi'_{i}=(1,{\eta'_{i}}_{2}{\eta'_{i}}_{1})^{T}$ and
$\xi_{i}^{*}=(1,\eta_{i1}^{*}\eta_{i2}^{*})$. We first integrate out $\eta'_{i}$, then Eq. (\ref{rdrho}) becomes
\begin{eqnarray}
&&\rho(\eta_{f}^{*},{\eta'_f};t)=A(t)e^{\eta_{f}^{*}J_{2}(t)\eta'_{f}}\int d\eta_{i}^{*}d\eta_{i}e^{-\eta_{i}^{*}\eta_{i}+\eta_{f}^{*}J_{1}(t)\eta_{i}}  \notag\\
&&\times [\xi_{i}^{*}\rho_{{1}}\tilde{\zeta}
+\eta_{i}^{*}\rho_{{2}}J_{1}^{\dag}(t)\eta'_{f}
+\eta_{i}^{*}\rho_{{2}}J_{3}(t)\eta_{i}],
\end{eqnarray}
where $\tilde{\zeta}=(1,\zeta_{2}\zeta_{1})^{T}$ and the vector $\zeta=J_{1}^{\dag}(t)\eta'_{f}+J_{3}(t)\eta_{i}$.
Similarly we can apply the same procedure on $\eta_{i}$, and after integrating out all the initial degrees of freedom,
the Q symbol simplifies to
\begin{eqnarray}
\rho(\eta_{f}^{*},{\eta'_f};t)&=&\xi_{f}^{*}\rho_{{1}}^{f}\xi'_{f}
+\eta_{f}^{*}\rho_{{2}}^{f}\eta'_{f},
\label{finalrho}
\end{eqnarray}
where the coefficient matrix $\rho_{{1}}^{f}$ and $\rho_{{2}}^{f}$ are given by
\begin{subequations}
\begin{eqnarray}
\rho^{f}_{{1}}&=&A(t)\tilde{J}_{1}(\rho_{{1}}+[(\rho_{{1}})_{2,2}\det J_{3}-\mathrm{Tr}(\rho_{{2}}J_{3})]\sigma_{+}\sigma_{-})\tilde{J}_{1}^{\dagger} \notag\\
&&+A(t)[\mathrm{Tr}(\sigma_{y}J_{2}^{T}\sigma_{y}J_{1}\rho_{{2}}J_{1}^{\dagger}) \notag\\
&&-(\rho_{{1}})_{2,2}\mathrm{Tr}(\sigma_{y}J_{2}^{T}\sigma_{y}J_{1}\sigma_{y}J_{3}^{T}\sigma_{y}J_{1}^{\dagger})\notag\\
&&+((\rho_{{1}})_{1,1}-\mathrm{Tr}(\rho_{{2}}J_{3}) \notag\\
&&+(\rho_{{1}})_{2,2}\det J_{3})\det J_{2}]\sigma_{-}\sigma_{+}, \\
\rho^{f}_{{2}}&=&A(t)J_{1}(\rho_{{2}}-
(\rho_{{1}})_{2,2}\sigma_{y}J_{3}^{T}\sigma_{y}){J}_{1}^{\dagger} \notag\\
&&+A(t)[(\rho_{{1}})_{1,1}+(\rho_{{1}})_{2,2}\det J_{3}-\mathrm{Tr}(\rho_{{2}}J_{3})]J_{2},
\end{eqnarray}
\label{finalcof}
\end{subequations}
where
$\tilde{J}_{i}=\left(\begin{array}{cc}
               1 & 0 \\
               0 & \det J_{i}(t)  \\
\end{array}\right)$ and $J_{i}\equiv J_{i}(t)$ for $i=1,2,3$. The Pauli matrix and the ladder operators are defined as
\begin{eqnarray}
\sigma_{y}=\left(\begin{array}{cc}
               0 & -i \\
               i & 0 \\
\end{array}\right),~
\sigma_{+}=\left(\begin{array}{cc}
               0 & 1 \\
               0 & 0 \\
\end{array}\right),~
\sigma_{-}=\left(\begin{array}{cc}
               0 & 0 \\
               1 & 0 \\
\end{array}\right).~
\end{eqnarray}
In the fermionic coherent states representation, the $4\times4$ coefficient matrix $\left(\begin{array}{cc}
               \rho^{f}_{{1}} & 0        \\
               0 & \rho^{f}_{{2}}        \\
\end{array}\right)$ completely determines the exact decoherence dynamics of the fermionic system.
Meanwhile the non-Markovian effect is fully manifested in the nonequilibirum Green's function $\mathcal{U}(t)$ and the
nonequilibrium thermal fluctuation correlation function $\mathcal{V}(t)$. In addition, the derivation of the reduced density matrix (\ref{finalrho}) is fully non-perturbative. It is valid for arbitrary time correlation functions $g(\tau)$ and
$\tilde{g}(\tau)$.

\subsection{Entanglement between fermionic modes}

The coefficient matrix Eq. (\ref{finalcof}) provide a direct way to explore the dynamics of fermionic entanglement when an
appropriate definition of entanglement is given. However the description of entanglement in fermionic systems is more
complicated \cite{Friis87 022338,Banuls76 022311,Keyl51 023522} than that in systems consisted of distinguishable two-level
systems, due to the fact that fermions are indistinguishable and anticommutative. The usually adopted definition of entanglement depends on the tensor product structure of the state space of the composite system. It reflects the particle aspect of first quantization rather than the collective, global aspect of second quantization. When we consider fermionic systems, however, such a
structure of the Hilbert space is no longer available. We therefore adopt the following definition of entanglement for
fermionic systems \cite{Friis87 022338},
\begin{eqnarray}
\bar{E}(\rho)=\min\limits_{p_{n},||\Psi_{n}\rangle\rangle}\sum\limits_{n}p_{n}S(||\Psi_{n}\rangle\rangle).
\label{DEoF}
\end{eqnarray}
The double-lined Dirac notation $||.\rangle\rangle$ denote states in fermionic Fock space in which states with more
than two fermions can be antisymmetrically constructed. The minimum is taken over all pure state ensembles $||\Psi_{n}\rangle\rangle$ that realize the state $\rho=\sum\limits_{n}p_{n}||\Psi_{n}\rangle\rangle\langle\langle\Psi_{n}||$ with $\sum\limits_{n}p_{n}=1$. $S(||\Psi_{n}\rangle\rangle)$ is the von Neumann entropy, which is a function of the
eigenvalues of the reduced states $Tr_{A}(||\Psi_{n}\rangle\rangle\langle\langle\Psi_{n}||)$ or $Tr_{B}(||\Psi_{n}\rangle\rangle\langle\langle\Psi_{n}||)$. According to Ref. \cite{Friis87 022338}, a general state in
the n-mode fermionic Fock space can be written as
\begin{eqnarray}
||\Psi\rangle\rangle&=&f_{0}||0\rangle\rangle+\sum_{i=1}^{n}{f_{i}||1_{i}\rangle\rangle}  \notag\\
&&+\sum_{j,k}^{n}f_{j,k}{||1_{j}\rangle\rangle||1_{k}\rangle\rangle}+...,
\end{eqnarray}
where $1_{i}$ denote an excitation in the mode $i$, and the coefficients $f_{j,k}$ form an antisymmetric matrix.
In addition $||1_{j}\rangle\rangle=\hat{a}_{j}^{\dagger}||0\rangle\rangle$ and $||1_{j}\rangle\rangle||1_{k}\rangle\rangle=\hat{a}_{j}^{\dagger}\hat{a}_{k}^{\dagger}||0\rangle\rangle$.

Back to Eq. (\ref{finalrho}), we note that fermionic coherent state is defined as the eigenvector of the annihilation
operator \cite{Cahill59 1538,Anastopoulos62 033821,Zhang62 867,Shresta71 022109}
\begin{eqnarray}
\hat{a} _{j}|\mathbf{\eta}\rangle=\eta_{j}|\mathbf{\eta}\rangle,~~~
|\mathbf{\eta}\rangle\equiv\prod\limits_{j}\exp(\hat{a} _{j}^\dag\eta_{j})|0\rangle.
\end{eqnarray}
Thus it is easy to verify that $\langle\langle0||\eta\rangle=1$, $\langle\langle1_{j}||\eta\rangle=\langle\langle0||\hat{a}_{j}|\eta\rangle=\eta_{j}$ and
$\langle\langle1_{2}||\langle\langle1_{1}||\eta\rangle=\langle\langle0||\hat{a}_{2}\hat{a}_{1}|\eta\rangle=\eta_{2}\eta_{1}$.
This means the coefficient matrix (\ref{finalcof}) also describe the density matrix in fermionic Fock space, and the fermionic
Fock space and fermionic coherent state representation both provide an intrinsic descriptions of fermionic states. If we restrict
the fermionic entanglement of formation (EoF) Eq. (\ref{DEoF}) to the case that respect superselection rules, then as pointed out
in Ref. \cite{Caban38 L79}, the minimization over all states of two fermionic modes can indeed be carried out. Finally we find
the explicit formula for the fermionic EoF
\begin{eqnarray}
\bar{E}(\rho)=-\frac{1}{2}\sum_{\sigma={1},{2}}Tr(\rho^{f}_{\sigma})\mathcal{K}_{\sigma},
\label{E1}
\end{eqnarray}
if $(\rho^{f}_{\sigma})_{1,1}=(\rho^{f}_{\sigma})_{2,2}$ and $(\rho^{f}_{\sigma})_{1,2}=0$ then
$\mathcal{K}_{\sigma}=0$, otherwise
\begin{eqnarray}
\mathcal{K}_{\sigma}=
(1-\lambda_{\sigma})\log_{2}\frac{1-\lambda_{\sigma}}{2}+(1+\lambda_{\sigma})\log_{2}\frac{1+\lambda_{\sigma}}{2},
\label{E2}
\end{eqnarray}
where
\begin{eqnarray}
\lambda_{\sigma}=\frac{(\rho^{f}_{\sigma})_{1,1}-(\rho^{f}_{\sigma})_{2,2}}{\sqrt{[(\rho^{f}_{\sigma})_{1,1}-%
(\rho^{f}_{\sigma})_{2,2}]^{2}+4|(\rho^{f}_{\sigma})_{1,2}|^{2}}}.
\label{E3}
\end{eqnarray}

\subsection{Analysis of the Green's function}

In the following, we focus on the Green's function $\mathcal{U}(t)$, which can induce vastly different dissipations
and fluctuations through different forms of the spectral density.
According to Eq. (\ref{coJ}), (\ref{v}) and (\ref{finalcof}), the solution of the Dyson equation (\ref{u}) uniquely
determines the evolution of the system. A general solutions to such integrodifferential equation have been derived
recently, utilizing the modified Laplace transformation \cite{Zhang109 170402}. It is straightforward to apply these
results to the case of interacting fermions, we then find the dissipationless non-Markovian dynamics (in other word it
is referred to as a process of nonthermal stabilization, in which the system still maintains partially its initial
information, and not reach thermal equilibrium with the environment) exist only when
\begin{eqnarray}
\det \left[\omega\mathbf{I}-\mathbf{M}-\int\frac{d\omega'}{2\pi}\frac{\mathbf{J}(\omega')}{\omega-\omega'}\right]=
\mathbf{J}_{kl}(\omega)=0.~~
\label{cd1}
\end{eqnarray}
Mathematically, this means real roots of the above equation exist only in the frequency regions that all spectral density
vanishes. From the physical point of view, it can also be explained by the bound state that generated between the
system and its environment \cite{John50 1764,Lodahl430 654,Bellomo78 060302}. A bound state is actually a stationary state
with a vanishing decay rate during the time evolution. If such a bound state is formed, it will lead to a dissipationless
dynamics. To illustrate this point clearly, we solve the Schr\"{o}dinger equation
\begin{eqnarray}
H_{tot}||\Psi_{E}\rangle\rangle=E||\Psi_{E}\rangle\rangle.
\label{SchrEq}
\end{eqnarray}
For simplicity, the environment is assumed to be zero temperature, and only one excitation is presented in the system initially,
then
\begin{eqnarray}
||\Psi_{E}\rangle\rangle=\sum\limits_{j=1}^{2}\left(c_{j}\hat{a}_{j}^\dag||vac\rangle\rangle
+\sum\limits_{lk}d_{lk}\hat{b}_{lk}^\dag||vac\rangle\rangle
\right),
\end{eqnarray}
where $||vac\rangle\rangle$ represent the vacuum state of the total system. Substituting Eq. (\ref{Htot}) into Eq. (\ref{SchrEq}),
one finds
\begin{subequations}
\begin{eqnarray}
\epsilon_{1}c_{1}+G c_{2}+\sum\limits_{lk}V_{1l}^{k}d_{lk}&=&E c_{1},
\label{c1} \\
\epsilon_{2}c_{2}+G^{*} c_{1}+\sum\limits_{lk}V_{2l}^{k}d_{lk}&=&E c_{2},
\label{c2} \\
\epsilon_{lk}d_{lk}+V_{1l}^{k*}c_{1}+V_{2l}^{k*}c_{2}&=&E d_{lk}.
\label{dlk}
\end{eqnarray}
\end{subequations}
Solving Eq. (\ref{dlk}) and substituting the solution $d_{lk}$ back into Eq. (\ref{c1}-\ref{c2}), we have
\begin{eqnarray}
&&\left(\begin{array}{cc}
\epsilon_{1}+\sum\limits_{lk}\frac{|V_{1l}^{k}|^{2}}{E-\epsilon _{lk}} &
G+ \sum\limits_{lk}\frac{V_{1l}^{k*}V_{2l}^{k}}{E-\epsilon _{lk}}  \\
G^{*}+ \sum\limits_{lk}\frac{V_{2l}^{k*}V_{1l}^{k}}{E-\epsilon _{lk}} &
\epsilon_{2}+\sum\limits_{lk}\frac{|V_{2l}^{k}|^{2}}{E-\epsilon _{lk}} \\
\end{array} \right)
\left(\begin{array}{c}
c_{1} \\ c_{2}
\end{array}\right) \notag\\
&&=E
\left(\begin{array}{c}
c_{1} \\ c_{2}
\end{array}\right).
\label{eigenEq}
\end{eqnarray}
By solving the eigenvalue equations (\ref{eigenEq}), and introducing the spectral density
$\sum\limits_{l}J_{jm}(\omega)=2\pi\sum\limits_{lk} V_{jl}^{k*}V_{ml}^{k}\delta(\omega-\omega_{k})$,
one finds the bound state exists only when
\begin{eqnarray}
\det \left[\left(\begin{array}{cc}
\epsilon_{1}-E & G   \\
G^{*}  & \epsilon_{2}-E \\
\end{array} \right)+\sum\limits_{l}\int\frac{d\omega}{2\pi}\frac{J_{l}(\omega)}{E-\omega}\right]=0.
\label{cd2}
\end{eqnarray}
Not surprisingly, Eq. (\ref{cd1}) and Eq. (\ref{cd2}) are completely equivalent. Based on these results,
we will present an analytical treatment of fermionic entanglement in the following sections.


\section{Analytical and numerical illustrations}

\subsection{Maximal fermionic entanglement in finite temperature}

To be more specific, we consider a system of DQDs coupled to electrodes (L and R) with all spins polarized in both
the dots and the electrodes. The corresponding Hamiltonian can be described by Eq. (\ref{Htot}) if one ignore the Coulomb
electron-electron interaction inside the dots \cite{Gurvitz53 15932,Tu78 235311,Jin12 083013,Yang89 115411}.
In general, the spectral density of the electrodes take an energy-dependent Lorentzian form
\cite{Jin12 083013,Zhang109 170402}
\begin{eqnarray}
\mathbf{J}_{kl}(\omega)=\frac{\Gamma_{l}d_{l}^{2}}{(\omega-\mu_{l})^{2}+d_{l}^{2}}\delta_{kl},
\label{Jsp}
\end{eqnarray}
where $l=L(R)$ for the left (right) electrode, $\Gamma_{l}$ is the coupling strength between the dot and electrode $l$,
$d_{l}$ is the bandwidth of the effective reservoir spectrum, and $\mu_{l}$ is the Fermi surface of the electron reservoir.

The usual Lorentzian spectral density (\ref{Jsp}) has been widely used to describe a fermionic environment.
The frequency region of $\mathbf{J}(\omega)$ cover the entire real axis,
according to Eq. (\ref{cd1}), therefore $\mathcal{U}(t)$ will present nonexponential (or exponential) decay during evolution
and approach zero ultimately. When the DQDs reach thermal equilibrium with the electrodes, the
steady state thermal fluctuations are fully determined by $\mathcal{V}^{s}$ (superscript $s$ denotes the steady state).
Then the general solution (\ref{finalcof}) will evolve into a steady state with the simple form
\begin{subequations}
\begin{eqnarray}
\rho^{s}_{1}&=&\left(\begin{array}{cc}
               1+\det\mathcal{V}^{s}-\mathrm{Tr}\mathcal{V}^{s} & 0 \\
               0 & \det\mathcal{V}^{s} \\
\end{array}\right), \\
\rho^{s}_{2}&=&\left(\begin{array}{cc}
               \mathcal{V}_{11}^{s}-\det\mathcal{V}^{s} & \mathcal{V}_{12}^{s} \\
               {\mathcal{V}_{12}^{s}}^{*} & \mathcal{V}_{22}^{s}-\det\mathcal{V}^{s} \\
\end{array}\right).
\end{eqnarray}
\label{rhos}
\end{subequations}
This result indicates that the corresponding system dissipate into a thermal-like \cite{Xiong1311 1282} (or thermal) state, and
its initial state information is completely washed out by environment. The corresponding fermionic EoF also reduce to
\begin{eqnarray}
\bar{E}^{s}
=(\det\mathcal{V}^{s}-\frac{1}{2}\mathrm{Tr}\mathcal{V}^{s})\mathcal{K}_{2}^{s}.
\label{EoFs}
\end{eqnarray}
From Eq. (\ref{rhos}) and (\ref{EoFs}), approximately, we find a positive correlation between $\bar{E}^{s}$ and the ratio of
$\frac{\mathrm{Tr}\mathcal{V}^{s}}{\det\mathcal{V}^{s}}$, especially, when
$\mathcal{V}_{11}^{s}=\mathcal{V}_{22}^{s}=|\mathcal{V}_{12}^{s}|=\frac{1}{2}$, a maximal entanglement is
generated $\bar{E}^{s}=1$. Obviously, the states of the DQDs in this scenario should be the Bell states denoted by
$||\Psi^{s}_{\pm}\rangle\rangle=\frac{1}{\sqrt{2}}(\hat{a}_{1}^{\dagger}||vac\rangle\rangle
\pm\hat{a}_{2}^{\dagger}||vac\rangle\rangle)$. Now the critical problem is how to satisfy the above conditions of the
steady state Green's function $\mathcal{V}^{s}$. To answer this question, we solve (\ref{v}) in the long-time limit.

According to Ref. \cite{Zhang109 170402}, the solution of $\mathcal{U}(t)$ takes the form
\begin{eqnarray}
\mathcal{U}(t)
&=&\int\frac{d\omega}{2\pi}\frac{ie^{-i\omega t}}
{\omega\mathbf{I}-\mathbf{M}-\mathbf{\Delta}(\omega)+i\frac{\mathbf{J}(\omega)}{2}}  \notag \\
&=&\sum\limits_{j}\mathcal{Z}_{j}e^{-ir_{j}t}, \label{Up}
\end{eqnarray}
where matrix $\mathbf{J}(\omega)$ is the spectral density, and the real part of the self-energy correction $\mathbf{\Delta}(\omega)$ is the principal value of the following integral
\begin{eqnarray}
\mathbf{\Delta}_{kl}(\omega)=\int\frac{d\omega'}{2\pi}\frac{\mathbf{J}_{kl}(\omega')}{\omega-\omega'}
=\frac{\Gamma_{l}d_{l}(\omega-\mu_{l})}{2d_{l}^{2}+2(\omega-\mu_{l})^{2}}\delta_{kl}.
\end{eqnarray}
The poles $r_{j}$ in (\ref{Up}) are located in the lower-half complex plane with the corresponding residues
$\mathcal{Z}_{j}$. In view of the solution (\ref{Up}), we then find the steady state Green's function
$\mathcal{V}^{s}(t\rightarrow\infty)=\int \mathbf{v}(\omega)d\omega$ with
\begin{eqnarray}
\mathbf{v}(\omega)=\frac{1}{2\pi}\sum\limits_{jk}\mathcal{Z}_{k} \frac{\mathbf{J}(\omega)\bar{\mathbf{n}}(\omega,T)}
{(\omega-r_{k})(\omega-r_{j}^{*})} \mathcal{Z}_{j}^{\dagger}.
\label{vs}
\end{eqnarray}
In the low temperature and low Fermi surface limit, when $\omega>\mu$ then $\bar{\mathbf{n}}(\omega,T)\rightarrow0$, otherwise when $\omega<\mu$, $\frac{\mathbf{J}(\omega)}{(\omega-r_{k})(\omega-r_{j}^{*})}<\frac{\mathbf{\Gamma}}{(\omega-r_{k})
(\omega-r_{j}^{*})}\ll1$. This means $\mathcal{V}^{s}$ approaches zero if the chemical potential
$\mu\ll-1$, which is consistent with the result in Ref. \cite{Xiong86 032107}. On the other hand if both of the chemical
potential $\mu_{l}>\epsilon_{l}$, the electrons tend to flow into the DQDs to fill the relatively lower energy levels of the
system, and the coherence of $\mathcal{V}^{s}$ may be destroyed in this process. Since the entanglement depends on the
coherence of $\mathcal{V}^{s}$, it is necessary to keep the chemical potential close to the energy scale of the system as to
generate a large steady state entanglement $\bar{E}^{s}$.

\begin{figure}[tbp]
~~~~\includegraphics[width=7.6cm]{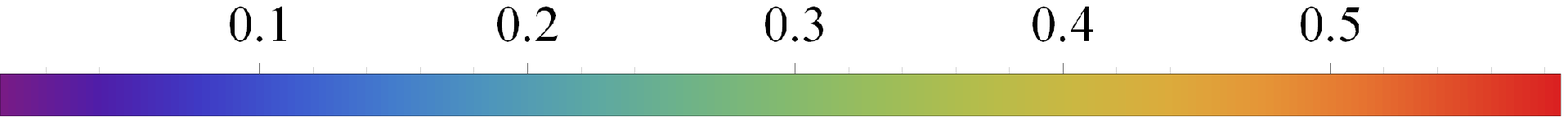}\\
\includegraphics[width=4.03cm]{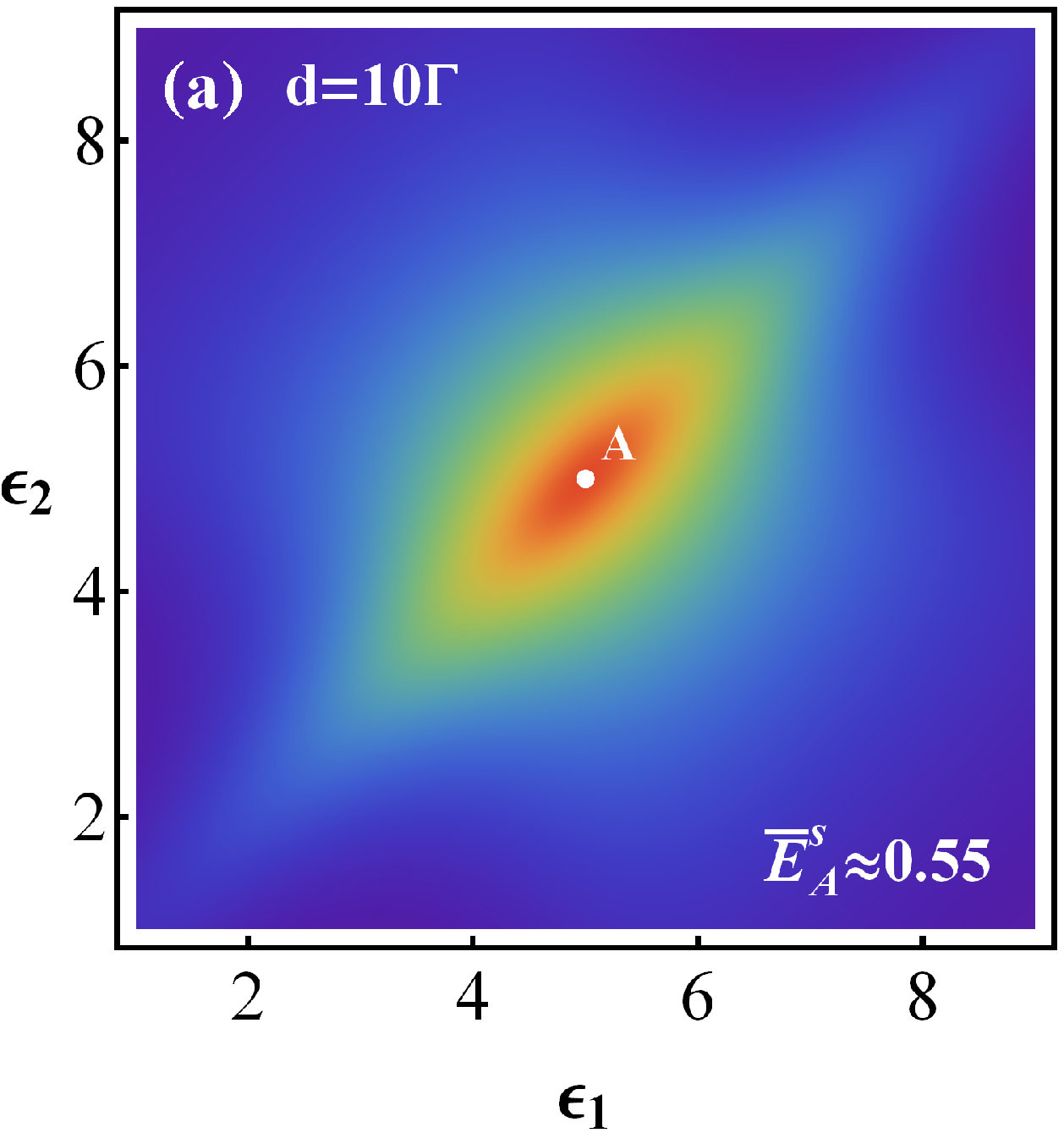}
\includegraphics[width=4.03cm]{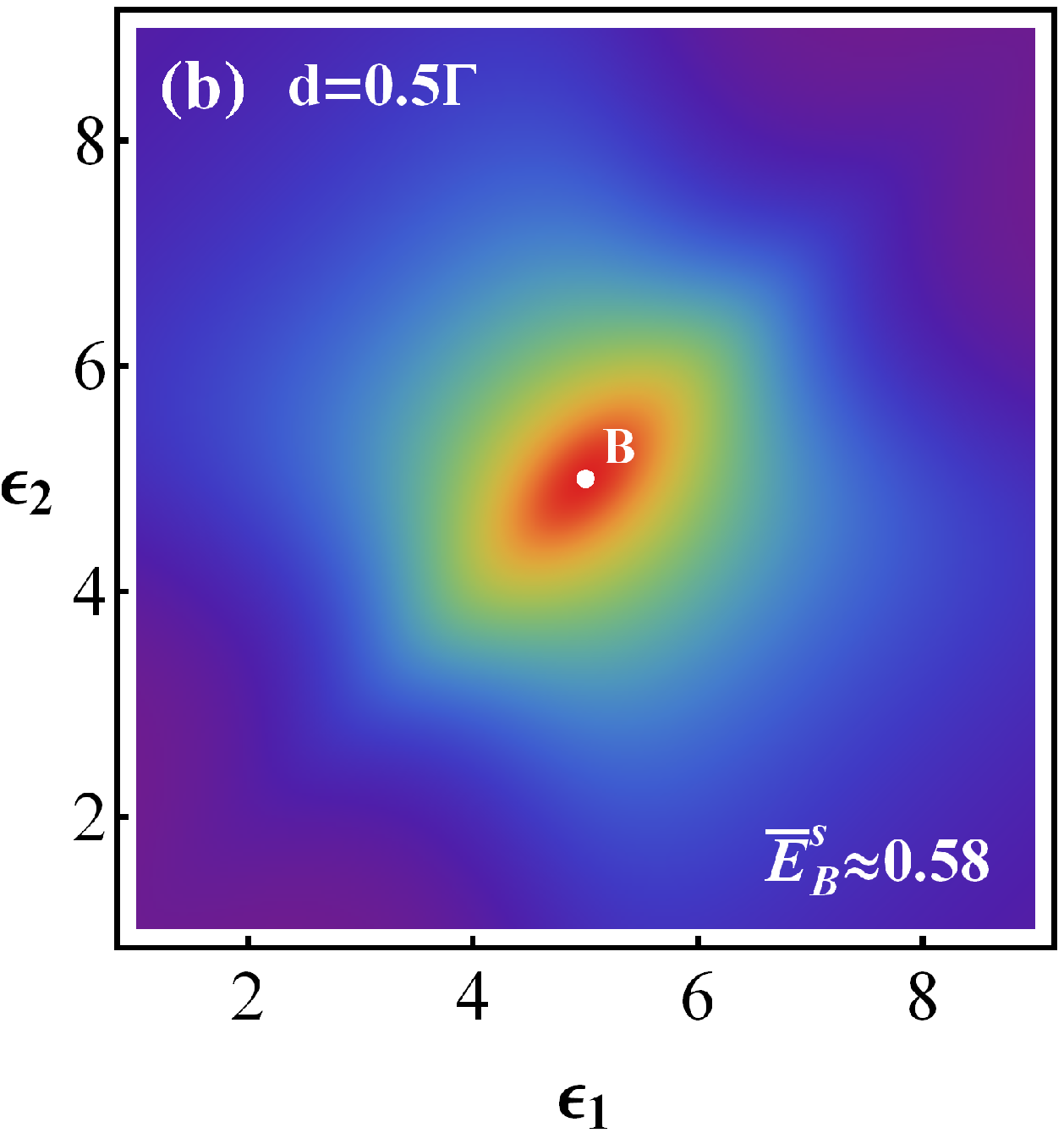}\\
\includegraphics[width=4cm]{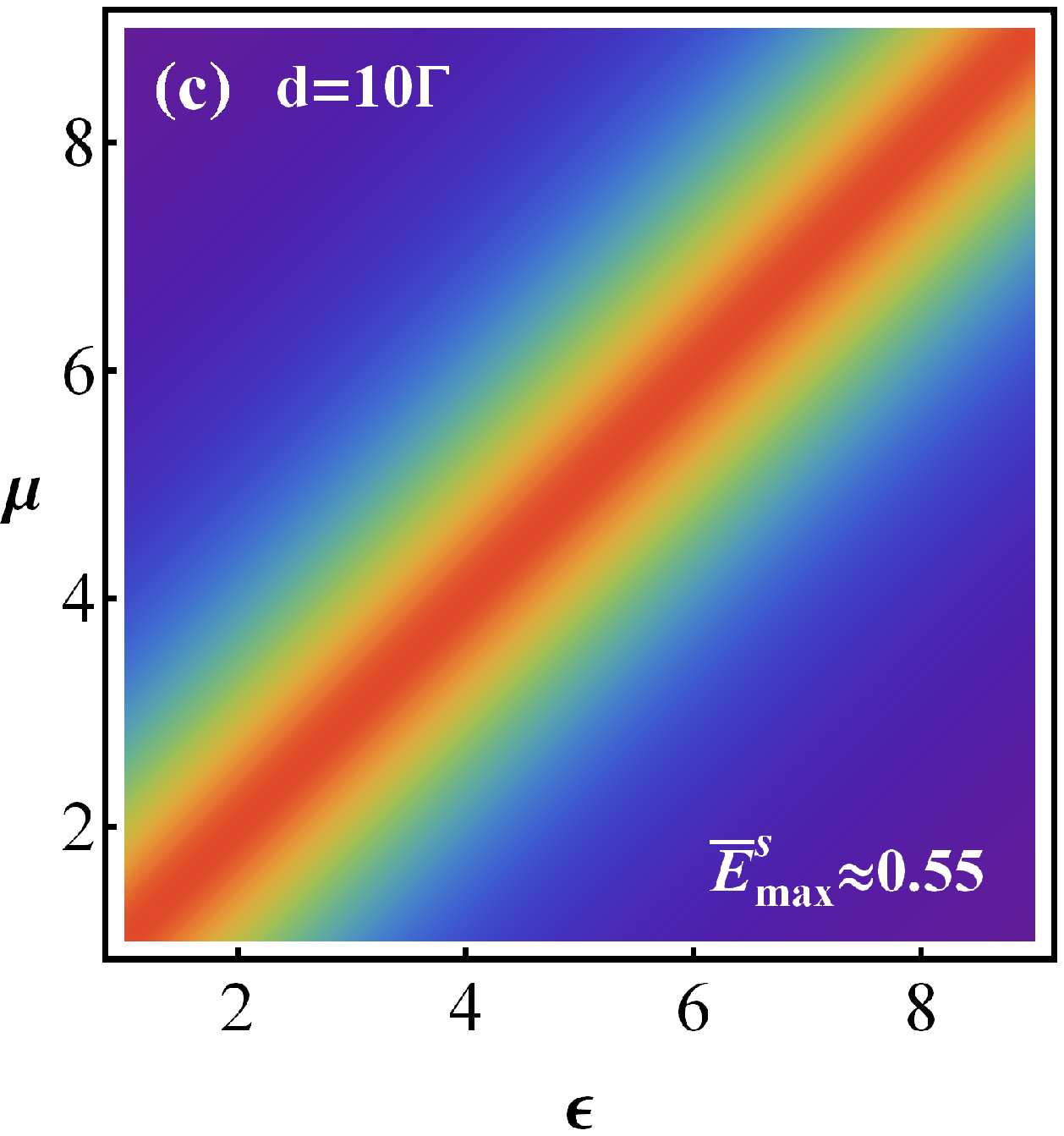}
\includegraphics[width=4cm]{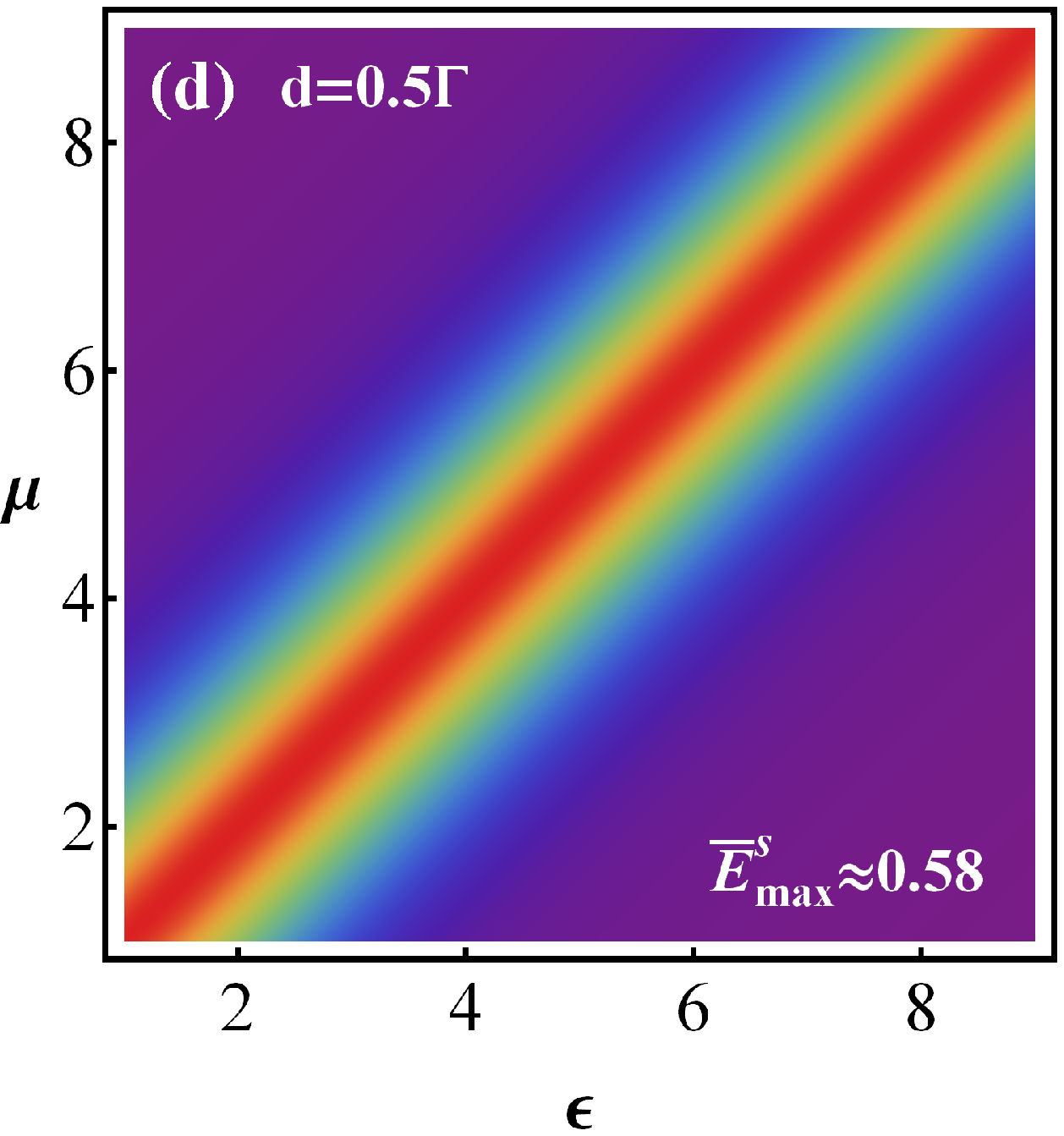}
\caption{(Color online) Density plot of steady state entanglement $\bar{E}^{s}$ versus the on-site energy $\epsilon_{l}$ and
the chemical potential $\mu_{l}$. In (a) and (b), we keep $\mu_{1}=\mu_{2}=5\Gamma$, but in (c) and (d), we take the
complete symmetric condition where $\epsilon_{1}=\epsilon_{2}=\epsilon$ and $\mu_{1}=\mu_{2}=\mu$. The other parameters are
$\Gamma_{1}=\Gamma_{2}=\Gamma$, $k_{B}T_{1}=k_{B}T_{2}=0.5\Gamma$ and $G=0.5\Gamma$.}
\label{f1}
\end{figure}

Using the solution of Eq. (\ref{vs}), combined with Eq. (\ref{Up}), we can easily calculate the steady state entanglement
$\bar{E}^{s}$ in Eq. (\ref{EoFs}). The result is plotted in Fig. \ref{f1} for the cases of $d=10\Gamma$
(weakly non-Markovian case) and $d=0.5\Gamma$ (strongly non-Markovian case). We assume the dot-environment interaction
is stronger than the tunnel coupling (i.e., $G<\Gamma$), in order to weaken the influences of the direct interaction
between fermionic modes and make a clarity analysis of the non-Markovian memory effects that impact on the system dynamics.
In Fig. \ref{f1}, we analyze two conditions where a symmetric chemical potential $\mu=5\Gamma$ and a complete symmetric
condition are considered (corresponding to Fig. \ref{f1}a-b and Fig. \ref{f1}c-d respectively). Comparing with different
values of bandwidth $d$, we find that the non-Markovian memory effect has made contribution to the increase of the maximum
value $\bar{E}^{s}_{max}$(points A and B stand for the maximum value of steady state entanglement). This fact is also well
supported by the inset of Fig. 2a. In addition, our numerical calculations show that this result is still valid for different
values of $G$. The main feature of Fig. \ref{f1}, on the other hand, is that the maximum $\bar{E}^{s}_{max}$ appears only in the
parametric region where a symmetric condition $\mu=\epsilon$ is satisfied. This result is consistent with the result discussed
above, which also indicates the symmetry of the system and environment can have significant impact on the dynamics of
entanglement. In the following, we thus go to the coherent manipulation regime, where the two QDs are set to be resonance,
and the electrodes are symmetric.

\begin{figure}[tbp]
\includegraphics[width=7cm]{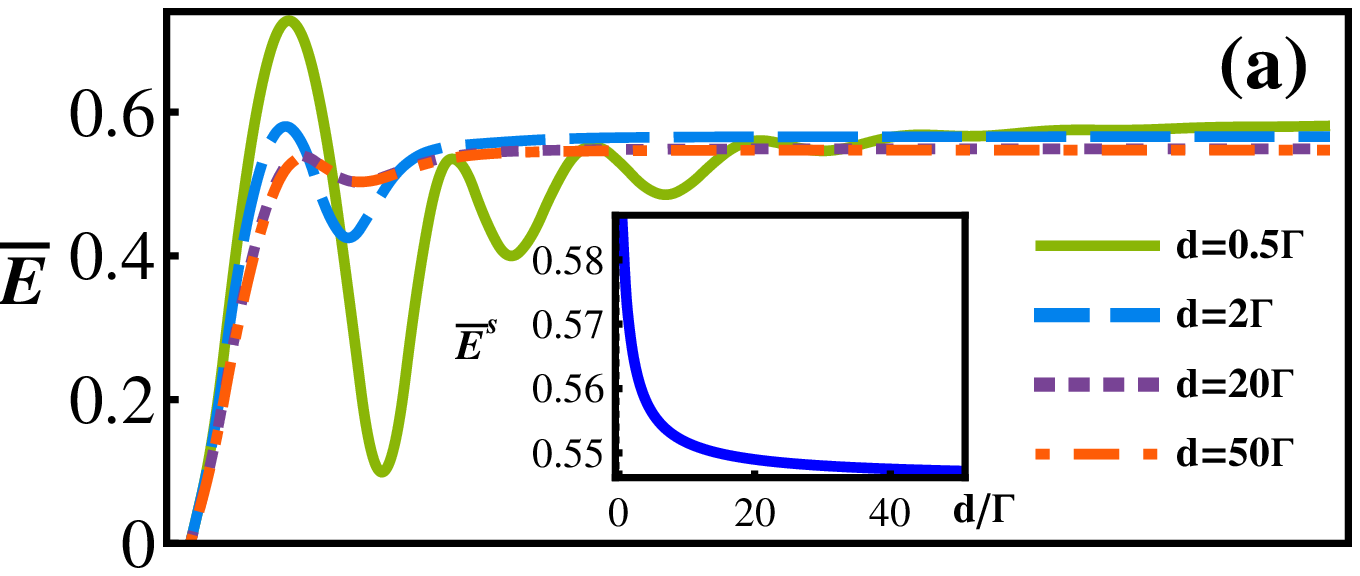}\\
\includegraphics[width=7cm]{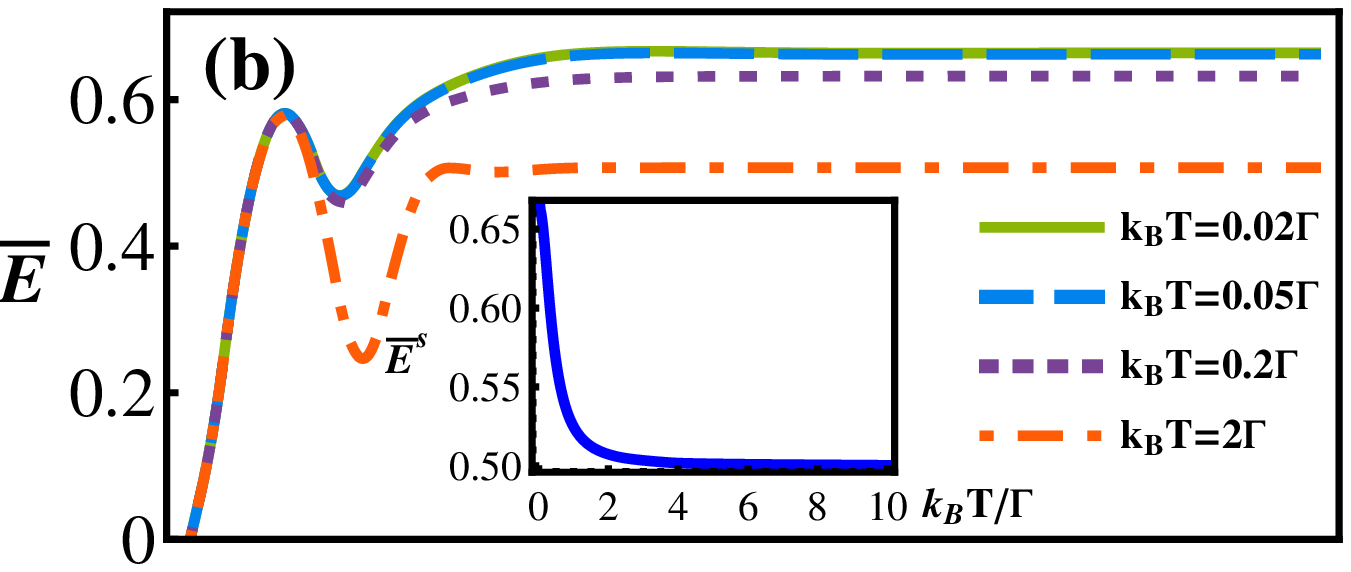}\\
\includegraphics[width=7cm]{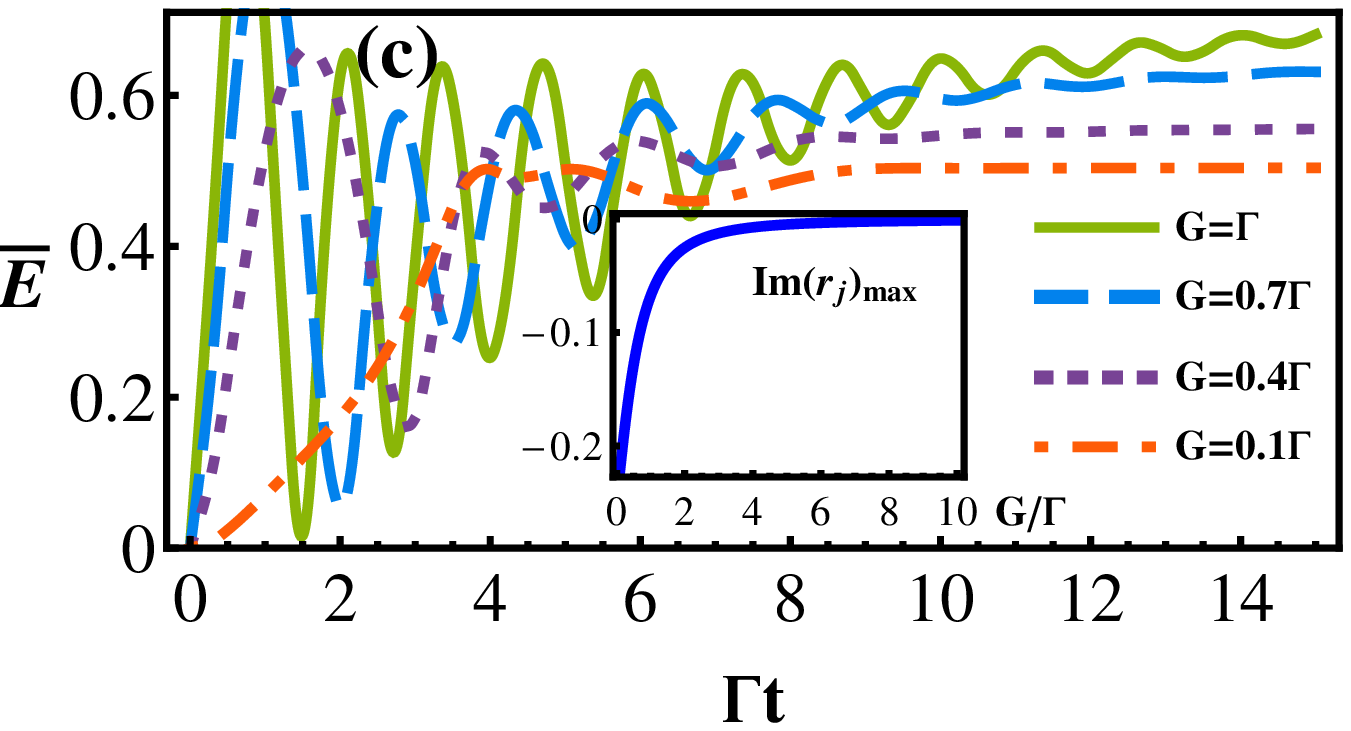}
\caption{(Color online) Entanglement evolution of the DQDs with Lorentzian spectral density,
where (a): $k_{B}T=G=0.5\Gamma$, (b): $d=2\Gamma$, $G=0.5\Gamma$ and (c):
$k_{B}T=d=0.5\Gamma$. Insets show the asymptotic dynamics of $\bar{E}$ with different bandwidths and temperatures.
The initial state is $\hat{a}_{1}^{\dag}||vac\rangle\rangle$. The other parameter are $\epsilon=\mu=2\Gamma$. Here we
consider the complete symmetric condition, so all the subscripts are omitted.
}
\label{f2}
\end{figure}

In Fig. \ref{f2}, we explore the fermionic EoF via the time dependent transient dynamics as well as the asymptotic limit
with different values of the bandwidth, temperature and interdot tunnel coupling. For the bandwidth $d\gg2\Gamma$,
$\bar{E}$ monotonically increases with time (neglecting the small oscillation) in a short-time scale, which is a result
similar to the Markovian dynamics. When the bandwidth $d<2\Gamma$, $\bar{E}$ approaches a steady value in the long-time
limit after several rounds of oscillation, which indicates a significant backaction effect that inducing the short-time
oscillation of the entanglement. The asymptotic value $\bar{E}^{s}$, as depicted in the inset of Fig. \ref{f2}a, shows
a small variation versus the bandwidth. In the present case, this means the bandwidth only has an important influence on
the transient dynamics, and becomes less important in the asymptotic limit. To show the effect of temperature on the
entanglement dynamics, we plot the fermionic EoF in Fig. \ref{f2}b at various temperatures. Note that $T$ has negligible
impact on the transient dynamics $t\lesssim1/\Gamma$, but manifests its action in the long-time scale. According to the
inset, one can also see that the steady state entanglement is sensitive to the temperature, and the magnitude of $\bar{E}^{s}$
decreases with the rising temperature. In the weak interdot tunnel coupling region $G\lesssim0.1\Gamma$ shown in Fig.
\ref{f2}c, $\bar{E}$ monotonically increases with time then approaches a steady value. Further increasing of $G$, a rapid
oscillation of $\bar{E}$ is observed in a short-time scale due to the enhancement of energy exchange of the DQDs.
The asymptotic behavior of $\bar{E}$ suggest that $\bar{E}^{s}$ is sensitive to $G$, and it increases with the increasing of
$G$.

In addition, Fig. 2c shows that the required time to reach the steady-state is getting longer when $\frac{G}{\Gamma}$ increases.
Physically, this phenomenon can be expected from the total Hamiltonian of the system, i.e., Eqs. (\ref{Htot}). The inter-dot
coupling $G$ plays a role in exchanging the population of the two fermionic modes. When the system evolves, some oscillations
may present in the system dynamics if this parameter is strong enough (as it is show in Fig. 2c, $G\approx\Gamma$). On the other
hand, the asymptotic behavior of the system dynamics can be attributed to the dot-environment interaction $\Gamma$. It reflects
the mutual time scale that arising from the environment. The competition between these two parameters determines the required
time that the system need to reach the steady-state.
Mathematically, the system decoherence dynamics is primarily determined by the retarded Green's function $\mathcal{U}(t)$,
and it approaches zero ultimately because the poles in Eq. (\ref{Up}) should satisfy $Im~r_{j}<0$. Thus the asymptotic dynamics
is determined by $\mathcal{Z}_{j}e^{-ir_{j}t}$ with the biggest imaginary part of $r_{j}$. One can expect that $\max\{Im~r_{j}\}$ increases with the increasing of $\frac{G}{\Gamma}$. This is proved by our numerical result shows in the inset of Fig. 2c.

\begin{figure}[tbp]
\includegraphics[width=7cm]{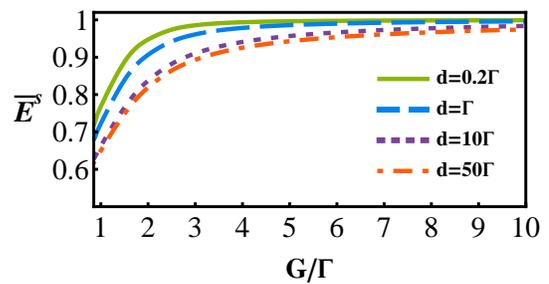}
\caption{(Color online) Steady state entanglement $\bar{E}^{s}$ versus the interdot tunnel coupling $G$. The DQDs
is set to be resonance with $\epsilon=\mu=2\Gamma$ and the temperature $k_{B}T=0.5\Gamma$. }
\label{f3}
\end{figure}

Based on the analytical and numerical analysis above, we then find the main factors that restricting the steady state
entanglement, which can be summarized as the symmetric condition, the interdot tunnel coupling and the initial temperature.
In Fig. \ref{f3}, as we expected, $\bar{E}^{s}$ reaches the maximum value in the strong interdot tunnel coupling region.
At this moment, the bandwidth also plays an important role. Compering with Fig. \ref{f2}a, we find that the non-Markovian
effect becomes significant when the interdot tunnel coupling is approximately in the region $\Gamma\lesssim G\lesssim4\Gamma$.
In the system under consideration, $G$ is a critical element in the generation of entanglement because it is the direct
interaction between the two QDs. If $G\ll\Gamma$, a very weak interdot tunnel coupling should be the main limiting
factor in entanglement generation. On the contrary, the system-environment interaction becomes negligible when $G\gg\Gamma$,
thus the non-Markovian effect manifests itself only in the region $G$ is comparable to $\Gamma$.

\subsection{Comparison to the Markovian dynamics}

In the previous subsections, we have shown the generation of maximal fermionic entanglement for DQDs coupled to
non-Markovian reservoirs. We found that the contribution of non-Markovian effect to the generation of entanglement
can not be ignored. For comparison, we should discuss the corresponding results in the Markovian approximation.

\begin{figure}[tbp]
~~~\includegraphics[width=7.69cm]{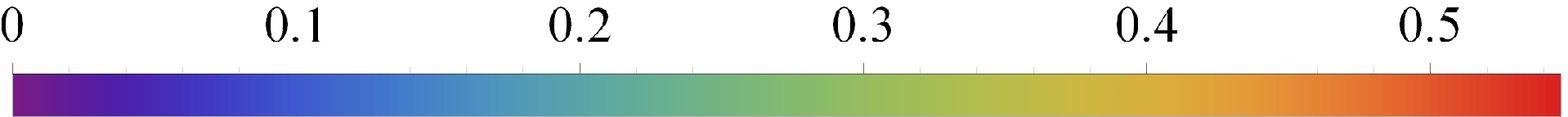}\\
\includegraphics[width=4.03cm]{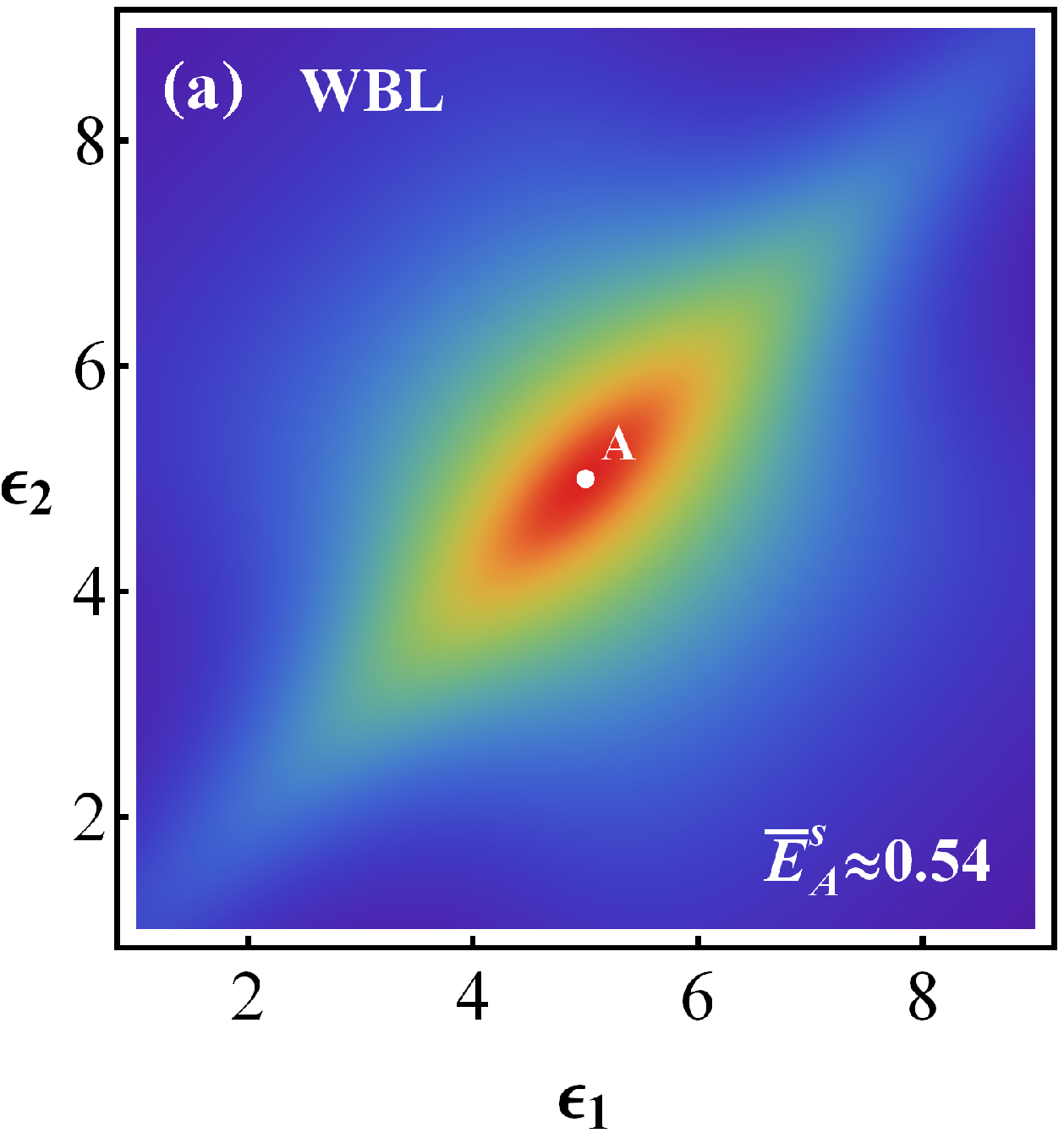}
\includegraphics[width=4.03cm]{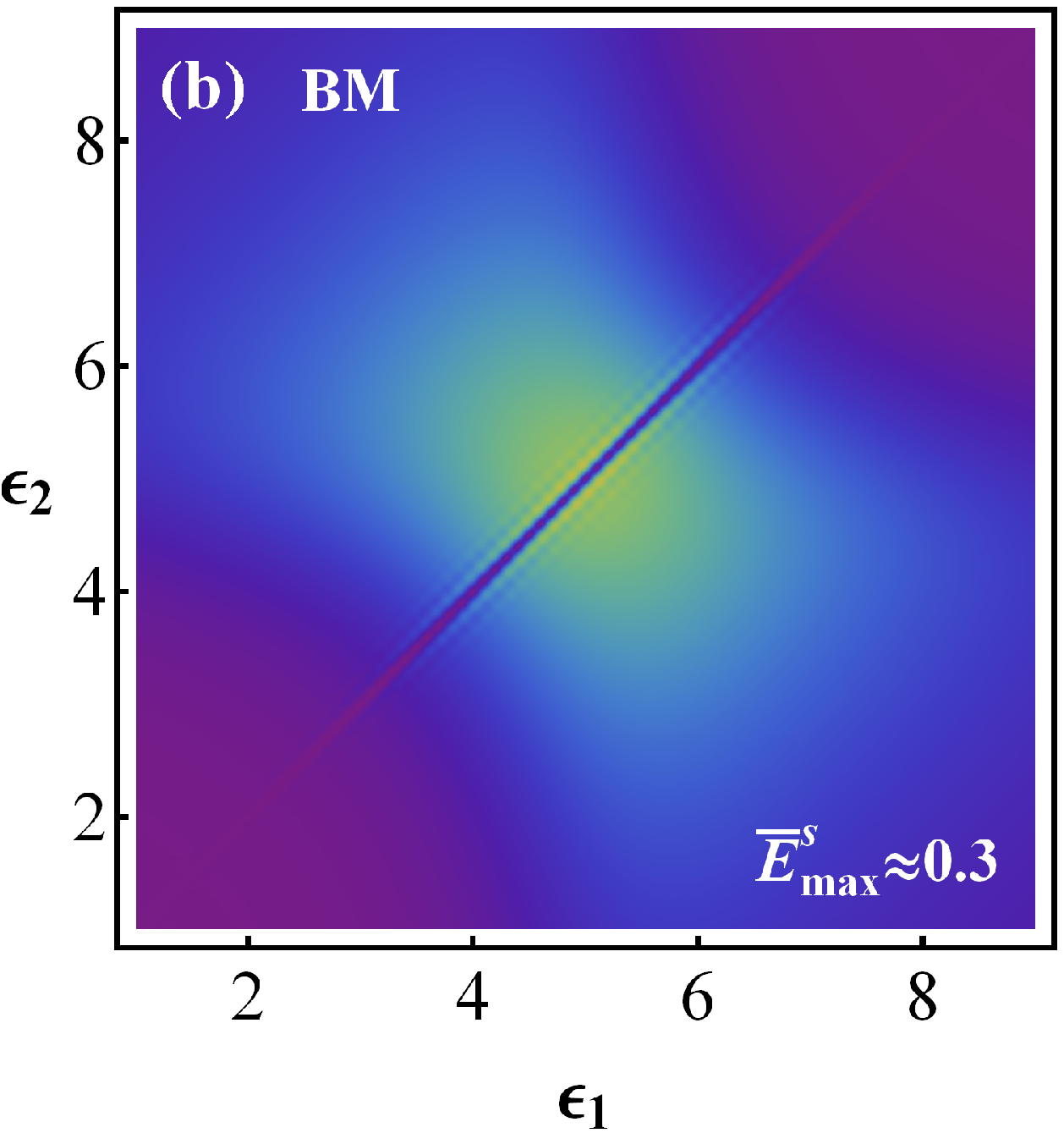}
\caption{(Color online) Density plot of steady state entanglement $\bar{E}^{s}$ versus the on-site energy $\epsilon_{l}$. The corresponding results of the wide-band limit (WBL) and Born-Markovian approximation (BM) is plotted in (a) and (b) respectively. We keep $\mu_{1}=\mu_{2}=5\Gamma$, and the other parameters are the same as that given in Fig. \ref{f1}.}
\label{f4}
\end{figure}

It is worth noting that one usually takes the wide-band limit ($d\rightarrow\infty$) of (\ref{Jsp}) to analyze various
quantum transport and quantum decoherence phenomena \cite{Imry2002}, which leads to the vanishment of the dominating
memory structure of $\mathcal{U}(t)$. In such a case, the spectrum of the reservoir takes a flat spectrum $J_{kl}(\omega)=\Gamma_{l}\delta_{kl}$, and the non-local time correlation functions are reduced to
\begin{eqnarray}
g_{_{WBL}}(\tau)&=&\Gamma\delta(\tau), \notag \\
\tilde{g}_{_{WBL}}(\tau)&=&\Gamma\int\frac{d\omega}{2\pi}\mathbf{\bar{n}}(\omega,T)e^{-i\omega\tau}. \notag
\end{eqnarray}
The correlation function $g_{_{WBL}}(\tau)$ is a delta function, which means that the memory structure of $\mathcal{U}_{WBL}(t)$
is completely washed out. However, for $\tilde{g}_{_{WBL}}(\tau)$, some memory effect still remains. The solutions of the
nonequilibirum Green's functions then reduced to
\begin{subequations}
\begin{eqnarray}
\mathcal{U}_{WBL}(t)&=&\exp\{-(i\mathbf{M}+\frac{\mathbf{\Gamma}}{2})t\},\\
\mathcal{V}_{WBL}(t)&=&\int\frac{d\omega}{2\pi} \frac{e^{-i\omega t}\mathbf{I}-e^{-(i\mathbf{M}
+\frac{\mathbf{\Gamma}}{2})t}}{\frac{\mathbf{\Gamma}}{2}+i(\mathbf{M}-\omega\mathbf{I})}
\mathbf{\bar{n}}(\omega,T)\mathbf{\Gamma}   \notag\\
&&\times\frac{e^{i\omega t}\mathbf{I}-e^{(i\mathbf{M}
-\frac{\mathbf{\Gamma}}{2})t}}{\frac{\mathbf{\Gamma}}{2}-i(\mathbf{M}-\omega\mathbf{I})}.~~~~~~
\end{eqnarray}
\end{subequations}
In the Born-Markovian dynamics, on the other hand, the coupling strength between the system and the environment is
very weak, and the characteristic correlation time of the environment $\tau_{E}=d^{-1}$ is sufficiently shorter than the
mutual time scale $\tau_{M}=\Gamma^{-1}$, i.e., $\tau_{E}\ll\tau_{M}$. In this case, it is believed that no memory effect
exists. Thus the differential equation of $\mathcal{U}_{BM}(t)$ and $\mathcal{V}_{BM}(t)$ at time $t$ should not depend
on it's past history. This requires the integro-differential equations
\begin{eqnarray}
\dot{\mathcal{U}}(\tau)+i \mathbf{M}\mathcal{U}(\tau)&+&\int_{0}^{\tau}d\tau'g(\tau-\tau')\mathcal{U}(\tau')=0, \notag\\
\dot{\mathcal{V}}(\tau)+i \mathbf{M}\mathcal{V}(\tau)&+&\int_{0}^{\tau}d\tau'g(\tau-\tau')\mathcal{V}(\tau') \notag\\
&=&\int_{0}^{t}d\tau'\tilde{g}(\tau-\tau')\overline{\mathcal{U}}(\tau'), \notag
\end{eqnarray}
reduced to differential equations \cite{Tu78 235311}. The corresponding correlation functions should be delta functions
\begin{eqnarray}
g_{_{BM}}(\tau)&=&\Gamma\delta(\tau), \notag \\
\tilde{g}_{_{BM}}(\tau)&=&\mathbf{\bar{n}}(\epsilon,T)\Gamma\delta(\tau). \notag
\end{eqnarray}
Here we can see that the differences between BWL and BM is that $\tilde{g}_{_{WBL}}(\tau)\neq\tilde{g}_{_{BM}}(\tau)$. It therefore leads to the different results of $\mathcal{V}_{WBL}(t)$ and $\mathcal{V}_{BM}(t)$. The true Markov limit is reached with the reduced Green's function
\begin{eqnarray}
\mathcal{V}_{BM}(t)=\int_{0}^{t}d\tau\mathcal{U}_{WBL}(t-\tau)\mathbf{\bar{n}}(\epsilon,T)\mathbf{\Gamma}
\mathcal{U}_{WBL}^{\dagger}(t-\tau).~~~
\label{VBM}
\end{eqnarray}

In Fig. \ref{f4}, we plot the steady state entanglement in the wide-band limit (a) and Born-Markovian approximation (b).
In the wide-band limit, the Dyson equation (\ref{u}) reduces to a time-convolutionless differential equation. Thus the
non-Markovian effect is completely lost in $\mathcal{U}_{WBL}$, but still manifested in $\mathcal{V}_{WBL}$, which explains
why Fig. \ref{f4}(a) manifests a similar behavior to Fig. \ref{f1}(a). In the Born-Markovian approximation, however, as shown
in Fig. \ref{f4}(b), the steady state entanglement vanishes on the purple line $\epsilon_{1}=\epsilon_{2}$,
but appears and becomes stronger in the neighboring areas around $\epsilon_{1}\approx\epsilon_{2}\approx5$. In other words,
the complete symmetric condition of the parameters has destructive effect on the steady state entanglement. This result is
quite surprising, because the Born-Markovian approximation lead to the conclusions that almost completely opposite of what we
obtained in the non-Markovian region. In this case, it is not a good approximation. Actually the unexpected result is a natural consequence of Eq. (\ref{VBM}), because $\mathcal{V}_{BM}(t)$ reduce to a diagonal matrix in the complete symmetric case, i.e., $\mathcal{V}_{BM}(t)=(\mathbf{I}-e^{-\mathbf{\Gamma}t})\mathbf{\bar{n}}(\epsilon,T)
=\frac{1-e^{-\Gamma t}}{e^{\beta(\epsilon-\mu)}+1}\mathbf{I}$. Finally, the DQDs reach thermal equilibrium with the electrodes,
$\mathcal{V}_{BM}^{s}=\mathbf{\bar{n}}(\epsilon,T)$, apparently, it is a standard thermal state.

\subsection{Lorentzian spectrum with a sharp cutoff}

\begin{figure}[tbp]
\includegraphics[width=4cm]{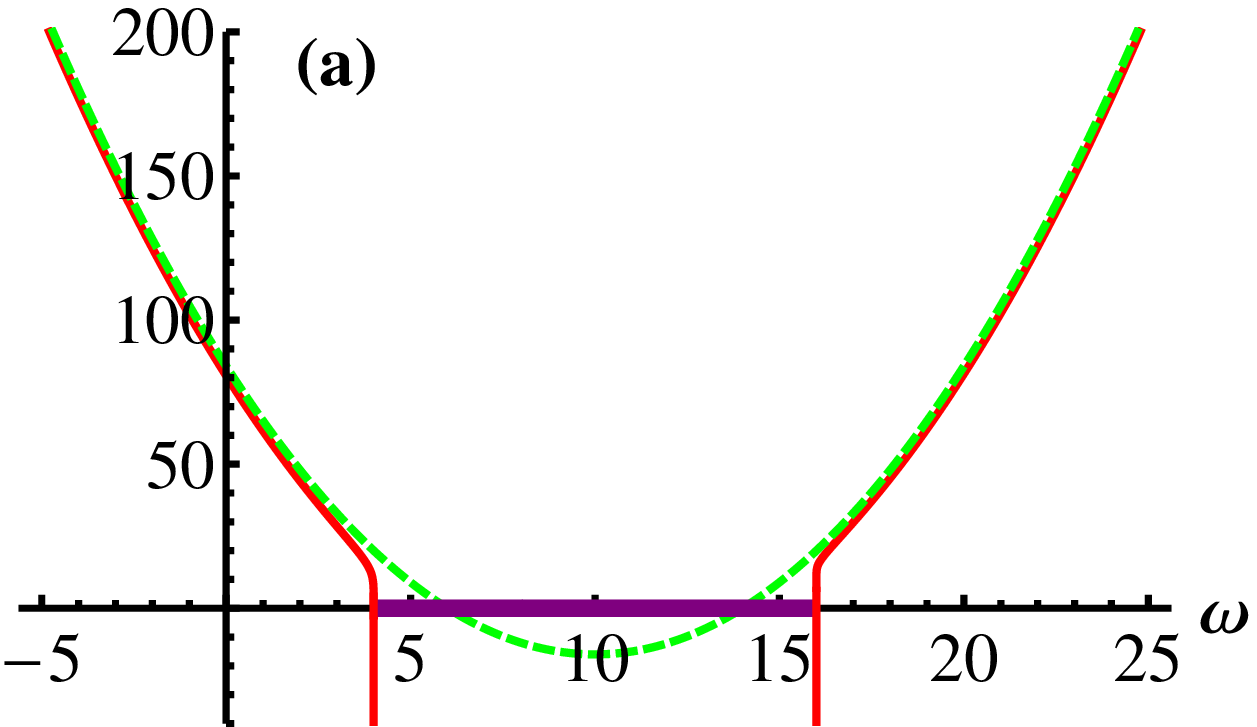}
\includegraphics[width=4cm]{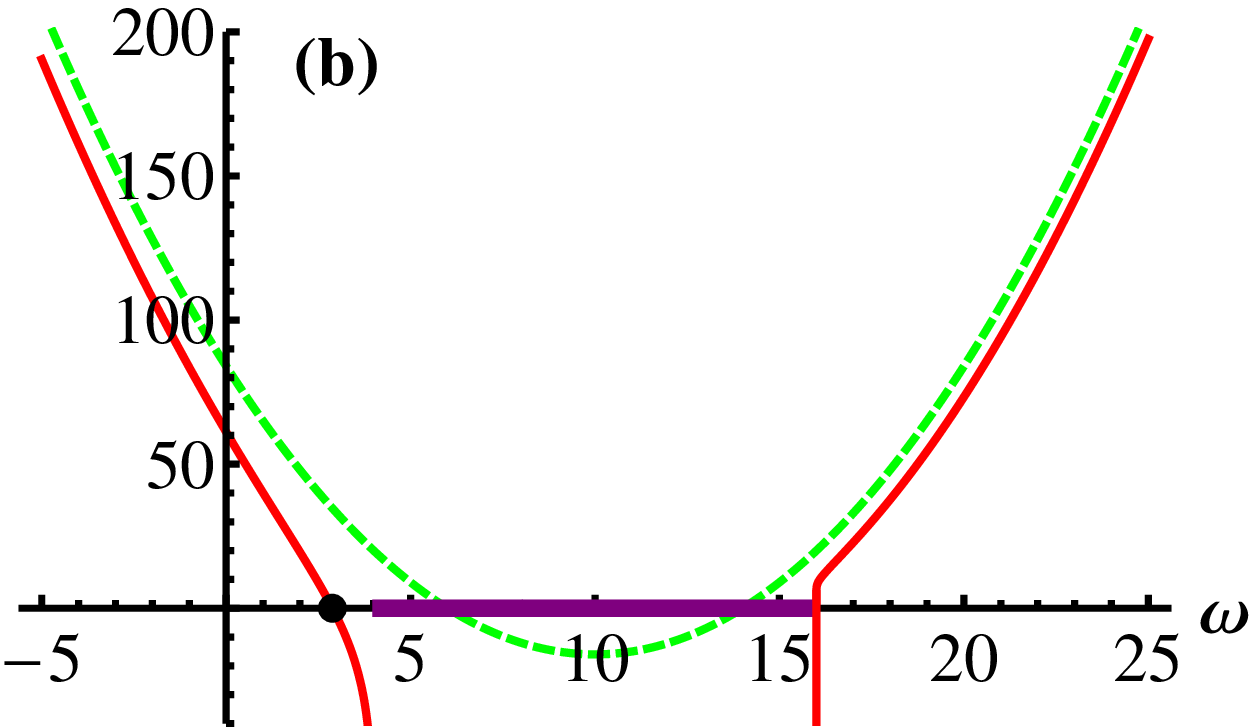}\\
\includegraphics[width=4cm]{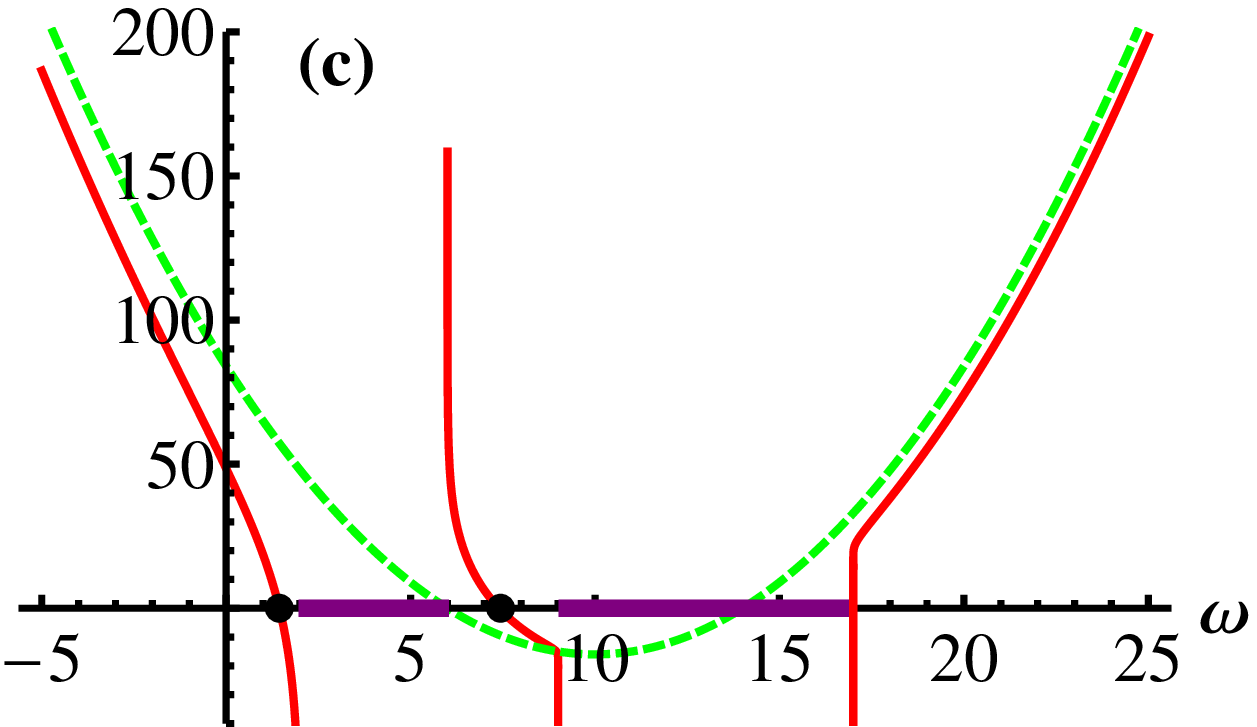}
\includegraphics[width=4cm]{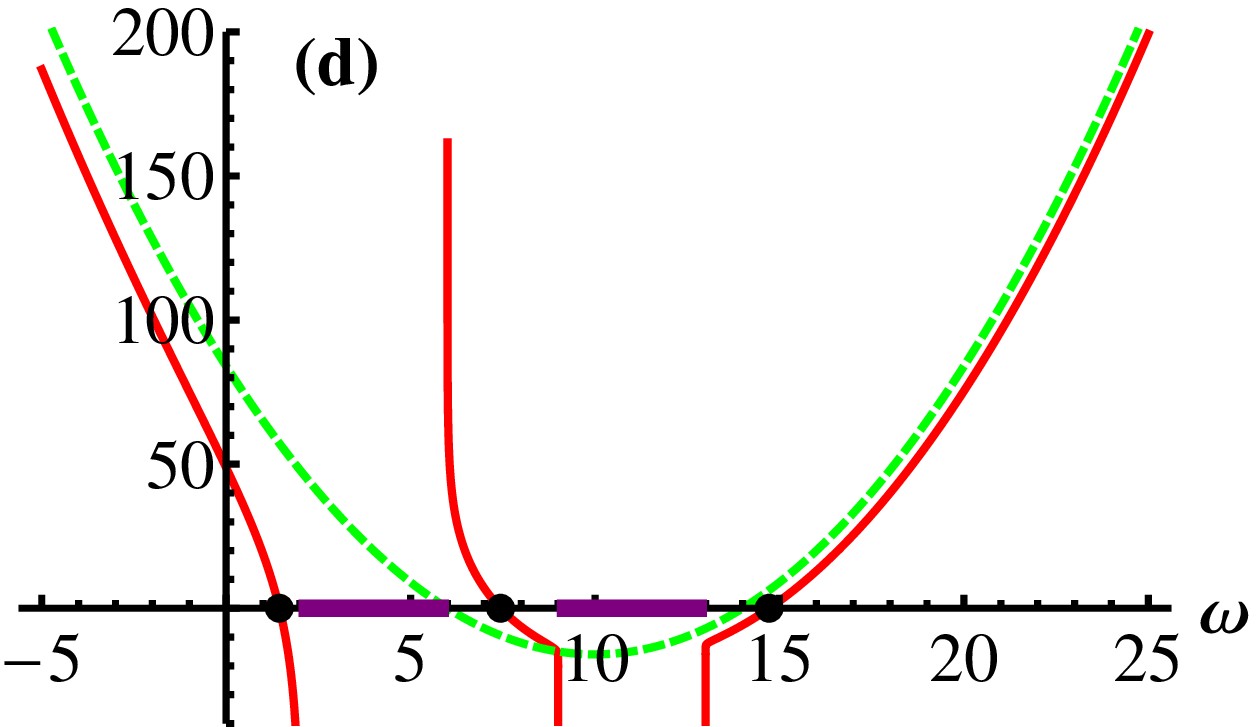}
\caption{(Color online) A schematic plots of the root structure of Eq. (\ref{cd1}), i.e., $\det\left[\omega\mathbf{I}-\mathbf{M}-\mathbf{\Sigma}(\omega)\right]=0$. The purple-shaded regimes on the real
axis $\omega$ correspond to the bandwidth region $\mathbf{J}(\omega)\neq0$. The green-dashed curve is the parabola
$\omega^{2}-(\epsilon_{1}+\epsilon_{2})\omega+\epsilon_{1}\epsilon_{2}-G^{2}$. }
\label{f5}
\end{figure}

The usual Lorentzian spectral density gives rise to nonexponential or exponential decay of the retarded Green's function
$\mathcal{U}(t)$. In this case, the initial state information is washed out completely, and the system may reach a steady
state with the environment. Now we discuss the scenario when the environmental density of states have a finite bandwidth.
This corresponds to add a sharp cutoff in the usual Lorentzian spectral density \cite{Zhang109 170402,Cai89 012128}
\begin{eqnarray}
\mathbf{J}_{kl}(\omega)=\frac{\Gamma_{l}d_{l}^{2}}{(\omega-\mu_{l})^{2}+d_{l}^{2}}\Theta(\Omega_{l}-|\omega-\mu_{l}|)\delta_{kl}.
\label{Jspcut}
\end{eqnarray}
When $\Omega\rightarrow\infty$, Eq. (\ref{Jspcut}) is reduced to the pure Lorentzian spectral density described by Eq.
(\ref{Jsp}) as we discussed in the previous section. The occurrence of the dissipationless non-Markovian dynamics requires
the spectrum to have at least one zero-value region. For a cutoff spectrum, it is found that the system-bath coupling
plays a critical role in the decoherence dynamics of a quantum system \cite{Cai89 012128}. One question naturally follows
for the cutoff Lorentzian-type spectrum (\ref{Jspcut}): How the spectral parameters affect the dynamic characteristic of
the system? To answer this question one need to understand the root structure of the criteria Eq. (\ref{cd1}).
Outside the band, the Laplace transformation of the self-energy takes the form
\begin{eqnarray}
&&\mathbf{\Sigma}_{kl}(\omega)=\int\frac{d\omega'}{2\pi}\frac{\mathbf{J}_{kl}(\omega')}{\omega-\omega'}
=\frac{\mathbf{J}_{kl}(\omega)}{2\pi}\notag\\
&&\times\left[
\log\left(\frac{\mu_{l}-\Omega_{l}-\omega}{\mu_{l}+\Omega_{l}-\omega}\right)
+\frac{2(\omega-\mu_{l})}{d_{l}}\arctan\left(\frac{\Omega_{l}}{d_{l}}\right)\right].~~~~~
\label{SEC}
\end{eqnarray}
Note that $\mathbf{\Sigma}(\omega)$ approaches to infinite on its boundary, and goes to zero asymptotically, i.e.,
$\mathbf{\Sigma}(\mu_{l}\pm\Omega_{l})=\pm\infty$ and $\mathbf{\Sigma}(\pm\infty)=0$. For large values of $\omega$,
the criteria $\det\left[\omega\mathbf{I}-\mathbf{M}-\mathbf{\Sigma}(\omega)\right]\approx\omega^{2}-(\epsilon_{1}
+\epsilon_{2})\omega+\epsilon_{1}\epsilon_{2}-G^{2}$. Thus, it is a parabola-kind curve with some gaps. One can change
the values of $\epsilon$ or $G$ to adjust the horizontal or vertical position of the curve. Meanwhile the central position
and the length of the gaps are determined by $\mu$ and $\Omega$. The corresponding schematic plots are shown in Fig.
\ref{f5}. There are basically zero or more effective roots for $\det\left[\omega\mathbf{I}-\mathbf{M}-\mathbf{\Sigma}(\omega)\right]=0$, which are denoted by the black points. We have
excluded the roots that are very close to the edge of the band gaps, because the long-time dissipationless dynamics is
depend on the residues of the Laplace transform of $\mathcal{U}(t)$ \cite{Zhang109 170402}, which are inverse proportion to $\det'\left[\omega\mathbf{I}-\mathbf{M}-\mathbf{\Sigma}(\omega)\right]$. Thus the roots near the edge have no contribution
to the dissipationless dynamics. The difference in the number of effective roots may induce vastly different dissipation
dynamics. There are three different relaxation processes \cite{Xiong1311 1282} for the spectral density (\ref{Jspcut}), the final states can be concluded as : thermal or thermal-like states for scenario (a); quantum memory states for
scenario (b); oscillating quantum memory states for scenario (c) and (d).

\section{Conclusion}

In conclusion, we have derived the exact solution of the reduced density matrix of a nanoelectronic system in finite-temperature non-Markovian reservoirs. The fermionic EoF is evaluated analytically by connecting the exact solution with an appropriate definition of fermionic entanglement in the fermionic Fock space. This provide us a potential way to extend the non-Markovian entanglement dynamics of distinguishable particles to the case of indistinguishable fermion systems. The system decoherence dynamics can be well described not only by the bound state between the system and its reservoirs, but also by the modified Laplace transformation of the Green's function. Our analysis shows that these two ways of description are completely equivalent for fermionic systems. Through our analytic and numerical calculations, we found that a maximal fermionic steady-state entanglement can be created in finite temperature non-Markovian environments. In the Born-Markovian approximation, the steady-state entanglement decreases and vanishes when the symmetric condition is considered. Our results pave the way to decoherence control of identical fermion systems, which deserves further investigation.


\begin{acknowledgments}
Acknowledgments: We would like to thank H. N. Xiong and P. Y. Lo for instructive discussions. This work is supported
by the NSF of China under Grant No. 11074028 and No. 11474044.
\end{acknowledgments}



\begin{thebibliography}{99}


\bibitem{Amico80 517} L. Amico, R. Fazio, A. Osterloh, and V. Vedral, Rev. Mod. Phys. 80, 517 (2008).

\bibitem{Horodecki81 865} R. Horodecki, P. Horodecki, M. Horodecki, and K. Horodecki, Rev. Mod. Phys. 81, 865 (2009).

\bibitem{Zyczkowski65 012101} K. \.{Z}yczkowski, P. Horodecki, M. Horodecki, and R. Horodecki, Phys. Rev. A 65, 012101 (2001).

\bibitem{Yu93 140404} T. Yu and J. H. Eberly, Phys. Rev. Lett. 93, 140404 (2004).

\bibitem{Sinaysky78 062301} I. Sinaysky, F. Petruccione, and D. Burgarth, Phys. Rev. A 78, 062301 (2008).


\bibitem{Wang15 103020} C. Wang and Q. H. Chen, New J. Phys. 15, 103020 (2013).

\bibitem{Benedetti87 052328} C. Benedetti, F. Buscemi, P. Bordone, and M. G. A. Paris, Phys. Rev. A 87, 052328 (2013).


\bibitem{Xu104 100502} J.-S. Xu, C.-F. Li, M. Gong, X.-B. Zou, C.-H. Shi, G. Chen, and
G.-C. Guo, Phys. Rev. Lett. 104, 100502 (2010).


\bibitem{Bellomo99 160502} B. Bellomo, R. Lo Franco, and G. Compagno, Phys. Rev. Lett. 99, 160502 (2007).

\bibitem{Maniscalco100 090503} S. Maniscalco, F. Francica, R. L. Zaffino, N. Lo Gullo, and F. Plastina,
Phys. Rev. Lett. 100, 090503 (2008).

\bibitem{Yang87 022312} W. Yang, J. H. An, C. Zhang, M. Feng, and C. Oh, Phys. Rev. A 87, 022312 (2013).

\bibitem{Tong81 052330} Q. J. Tong, J. H. An, H. G. Luo, and C. H. Oh, Phys. Rev. A 81, 052330 (2010).


\bibitem{John50 1764} S. John and T. Quang, Phys. Rev. A 50, 1764 (1994).

\bibitem{Lodahl430 654} P. Lodahl, A. F. van Driel, I. S. Nikolaev, A. Irman, K. Overgaag,
D. Vanmaekelbergh, and W. L. Vos, Nature (London) 430, 654 (2004).

\bibitem{Bellomo78 060302} B. Bellomo, R. LoFranco, S. Maniscalco, and G. Compagno,
Phys. Rev. A 78, 060302(R) (2008).


\bibitem{Zhang109 170402} W. M. Zhang, P. Y. Lo, H. N. Xiong, M. W. Y. Tu, and F. Nori,
Phys. Rev. Lett. 109, 170402 (2012).


\bibitem{Elzerman67 161308} J. M. Elzerman, R. Hanson, J. S. Greidanus, L. H. Willems van Beveren, S. De Franceschi,
L. M. K. Vandersypen, S. Tarucha, and L. P. Kouwenhoven, Phys. Rev. B 67, 161308(R) (2003).

\bibitem{Hayashi91 226804} T. Hayashi, T. Fujisawa, H. D. Cheong, Y. H. Jeong, and Y. Hirayama,
Phys. Rev. Lett. 91, 226804 (2003).


\bibitem{Recher63 165314} P. Recher, E. V. Sukhorukov, and D. Loss, Phys. Rev. B 63, 165314 (2001).

\bibitem{Legel76 085335} S. Legel, J. K\"{o}nig, G. Burkard, and G. Sch\"{o}n, Phys. Rev. B 76, 085335 (2007).

\bibitem{Erbe85 155127} B. Erbe and J. Schliemann, Phys. Rev. B 85, 155127 (2012).


\bibitem{Blattmann89 012327} R. Blattmann, H. J. Krenner, S. Kohler, and P. H\"{a}nggi, Phys. Rev. A 89, 012327 (2014).


\bibitem{Posazhennikova88 042302} A. Posazhennikova, R. Birmuske, M. Bruderer, and W. Belzig, Phys. Rev. A 88, 042302 (2013).

\bibitem{Hiltunen89 115322} T. Hiltunen and A. Harju, Phys. Rev. B 89, 115322 (2014).


\bibitem{Blaauboer95 160402} M. Blaauboer and D. P. DiVincenzo, Phys. Rev. Lett. 95, 160402 (2005).

\bibitem{Emary80 161309} C. Emary, Phys. Rev. B 80, 161309(R) (2009).

\bibitem{Borras84 033301} A. Borras and M. Blaauboer, Phys. Rev. B 84, 033301 (2011).


\bibitem{Thorwart72 235320} M. Thorwart, J. Eckel, and E. R. Mucciolo, Phys. Rev. B 72, 235320 (2005).

\bibitem{Liang72 245328} X. T. Liang, Phys. Rev. B 72, 245328 (2005).

\bibitem{Cao76 115301} X. Cao and H. Zheng, Phys. Rev. B 76, 115301 (2007).

\bibitem{Marcos83 125426} D. Marcos, C. Emary, T. Brandes, and R. Aguado, Phys. Rev. B 83, 125426 (2011).

\bibitem{Madsen106 233601} K. H. Madsen, S. Ates, T. Lund-Hansen, A. L\"{o}ffler, S. Reitzenstein, A. Forchel, and P. Lodahl,
Phys. Rev. Lett. 106, 233601 (2011).


\bibitem{Tu78 235311} M. W. Y. Tu and W. M. Zhang, Phys. Rev. B 78, 235311 (2008).

\bibitem{Jin12 083013} J. S. Jin, M. W. Y. Tu, W. M. Zhang, and Y. J. Yan, New J. Phys. 12, 083013 (2010).


\bibitem{Verch17 545} R. Verch and R. Werner, Rev. Math. Phys. 17, 545 (2005).


\bibitem{Li64 054302} Y. S. Li, B. Zeng, X. S. Liu, and G. L. Long, Phys. Rev. A 64, 054302 (2001).

\bibitem{Paskauskas64 042310} R. Paskauskas and L. You, Phys. Rev. A 64, 042310 (2001).

\bibitem{Schliemann64 022303} J. Schliemann, J. I. Cirac, M. Ku\'{a}, M. Lewenstein, and D. Loss, Phys. Rev. A 64, 022303 (2001).





\bibitem{Schliemann63 085311} J. Schliemann, D. Loss, and A. H.MacDonald, Phys. Rev. B 63, 085311 (2001).

\bibitem{Eckert299 88} K. Eckert, J. Schliemann, D. Bru$\beta$, and M. Lewenstein, Ann. Phys. 299, 88 (2002).

\bibitem{Shi67 024301} Y. Shi, Phys. Rev. A 67, 024301 (2003).

\bibitem{Wiseman91 097902} H. M. Wiseman and J. A. Vaccaro, Phys. Rev. Lett. 91, 097902 (2003).


\bibitem{Friis87 022338} N. Friis, A. R. Lee, and D. E. Bruschi, Phys. Rev. A 87, 022338 (2013).


\bibitem{Feynman24 118} R. P. Feynman and F. L. Vernon, Ann. Phys. (N.Y.) 24, 118 (1963).


\bibitem{Zhang62 867} W. M. Zhang, D. H. Feng, and R. Gilmore, Rev. Mod. Phys. 62, 867 (1990).

\bibitem{Cahill59 1538} K. E. Cahill and R. J. Glauber, Phys. Rev. A 59, 1538 (1999).


\bibitem{Berezin1966} F. A. Berezin, \textit{The Method of Second Quantization} (Academic Press, New York, 1966).




\bibitem{Banuls76 022311} M. C. Ba\~{n}uls, J. I. Cirac, and M. M.Wolf, Phys. Rev. A 76, 022311 (2007).

\bibitem{Keyl51 023522} M. Keyl and D. M. Schlingemann, J. Math. Phys. 51, 023522 (2010).


\bibitem{Anastopoulos62 033821} C. Anastopoulos and B. L. Hu, Phys. Rev. A 62, 033821 (2000).

\bibitem{Shresta71 022109} S. Shresta, C. Anastopoulos, A. Dragulescu, and B. L. Hu, Phys. Rev. A 71, 022109 (2005).


\bibitem{Caban38 L79} P. Caban, K. Podlaski, J. Rembieli\'{n}ski, K. A. Smoli\'{n}ksi, and
Z. Walczak, J. Phys. A: Math. Gen. 38, L79 (2005).


\bibitem{Gurvitz53 15932} S. A. Gurvitz and Ya. S. Prager, Phys. Rev. B 53, 15932 (1996).

\bibitem{Yang89 115411} P. Y. Yang, C. Y. Lin, and W. M. Zhang, Phys. Rev. B 89, 115411 (2014).

\bibitem{Xiong1311 1282} H. N. Xiong, P. Y. Lo, W. M. Zhang, F. Nori, and D. H. Feng, arXiv:1311.1282v1 (2014).

\bibitem{Xiong86 032107} H. N. Xiong, W. M. Zhang, M. W. Y. Tu, and D. Braun, Phys. Rev. A 86, 032107 (2012).

\bibitem{Imry2002} Y. Imry, \textit{Introduction to Mesoscopic Physics}, 2nd ed. (Oxford University Press, Oxford, 2002).

\bibitem{Cai89 012128} C. Y. Cai, L. P. Yang, and C. P. Sun, Phys. Rev. A 89, 012128 (2014).



\end{thebibliography}
\end{document}